\documentclass[letterpaper]{article} 
\usepackage{aaai25}  
\usepackage{xcolor}
\usepackage{caption}
\usepackage{subcaption}
\usepackage{times}  
\usepackage{helvet}  
\usepackage{courier}  
\usepackage[hyphens]{url}  
\usepackage{graphicx} 
\usepackage{natbib}  
\usepackage{caption} 
\usepackage{algorithm}
\usepackage{algorithmic}

\usepackage{newfloat}
\usepackage{listings}
\DeclareCaptionStyle{ruled}{labelfont=normalfont,labelsep=colon,strut=off} 
\floatstyle{ruled}
\newfloat{listing}{tb}{lst}{}
\floatname{listing}{Listing}
\newcommand{\answerYes}[1]{\textcolor{blue}{#1}} 
\newcommand{\answerNo}[1]{\textcolor{teal}{#1}} 
\newcommand{\answerNA}[1]{\textcolor{gray}{#1}} 

\newcommand{\revision}[1]{#1}

\title{Characterizing Online Activities Contributing to Suicide Mortality among Youth}
\author{
    Aparna Ananthasubramaniam\textsuperscript{\rm 1}, Elyse J. Thulin, Viktoryia Kalesnikava, Silas Falde, Jonathan Kertawidjaja, Lily Johns, Alejandro Rodríguez-Putnam, Emma Spring, Kara Zivin, Briana Mezuk 
}
\affiliations{
    University of Michigan\\

    \textsuperscript{\rm 1}Corresponding author: akananth@umich.edu
}

\usepackage{bibentry}

\begin{document}

\maketitle

\begin{figure*}[t]
\centering
\includegraphics[width=0.8\textwidth]{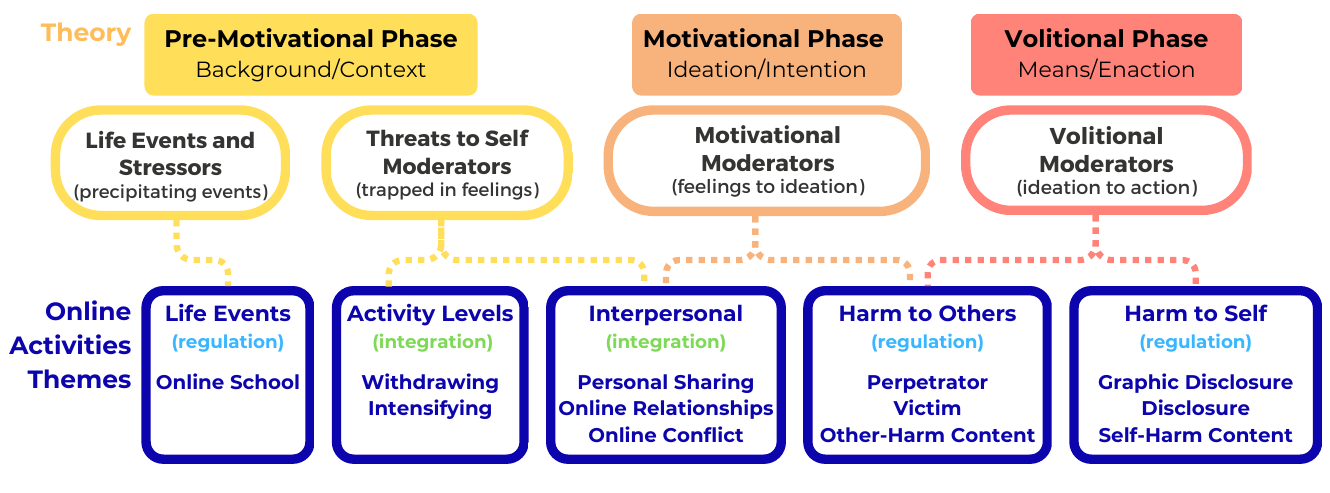}
\caption{This paper presents a theory-driven thematic analysis to characterize how online behaviors contribute to suicide risk among youth who died by suicide. \revision{Although the initial phases of the thematic analyses used open coding, themes were ultimately constructed to align with two theories of suicide: The first is the IMV theory that describes three phases of suicidal risk shown in yellow (pre-motivational phase), orange (motivational phase), and red (volitional phase) at the top of the figure. The second is Durkheim's theory of suicide, which describes suicide risk as the result of transgressions of the norms of integration (shown in light green in the bottom row of boxes) and regulation (shown in light blue). Using open coding and these two theories, we construct 12 themes related to the online activities described in NVDRS narratives, shown in the purple boxes at the bottom. These themes are related to a variety of topics, including harm to self, harm to others, interpersonal interactions, activity levels online, and life events. This figure depicts how each theme corresponds to a different IMV phase and set of Durkheim's norms. Detailed descriptions of the themes are given in Table~\ref{tab:themes}.} 
}
\label{fig:teaser}
\end{figure*}

\begin{abstract}
The recent rise in youth suicide highlights the urgent need to understand how online experiences contribute to this public health issue. Our mixed-methods approach responds to this challenge by developing a set of themes focused on risk factors for suicide mortality \revision{in online spaces among youth ages 10-24,} and a framework to model these themes at scale. \revision{Using 29,124 open text summaries of death investigations between 2013-2022,} we \revision{conducted a thematic analysis to} identify 12 \revision{types of online activities} that were considered by investigators or next of kin to be relevant in contextualizing a given suicide death. \revision{We then develop a zero-shot learning framework to model these 12 themes at scale, and analyze variation in these themes by decedent characteristics and over time. Our work uncovers several online activities related to harm to self, harm to others, interpersonal interactions, activity levels online, and life events, which correspond to different phases of suicide risk from two prominent suicide theories. We find an association between these themes and decedent characteristics like age, means of death, and interpersonal problems, and many themes became more prevalent during the 2020 COVID-19 lockdowns. While digital spaces have taken some steps to address expressions of suicidality online, our work illustrates the opportunities for developing interventions related to less explicit indicators of suicide risk by combining suicide theories with computational research.}
\end{abstract}

%

\section{Introduction}

{\color{red} \textbf{Content Warning:} This paper discusses suicide mortality among youth, including some graphic or explicit behaviors.}

Suicide is the third leading cause of death among youth ages 10 - 24 in the U.S.A. \citep{nimh2025suicide}. In fifteen years, youth suicide mortality has risen 62\%, from 6.8 deaths per 100,000 in 2007 to to 11.0 in 2021 \citep{nchs2023suicide}. The urgency to address suicide behaviors is matched by a growing set of concerns about youth mental health more generally \citep{brener2024overview}. In his 2021 advisory, the U.S. Surgeon General identified \revision{online spaces generally and} social media specifically as key factors in recent declines in youth mental health \citep{surgeon2021protecting}. By age 17, over 95\% of contemporary youth have access to a smartphone, and adolescents spend on average 4.8 hours a day on social media applications \citep{rothwell2023parenting}. Research links higher use of social media, longer screen times, and harmful online activities (i.e., cyberbullying, coercive sexting) with increased mental health challenges and youth suicide ideation \citep{thulin2024longitudinal}. 

Current research has largely focused on addressing this problem by classifying and understanding how to mitigate expressions of suicidal \textit{ideation and intention} online \citep{macrynikola2021does}. However, it is unclear which specific online behaviors are linked to youth suicide \textit{deaths}. \revision{Additionally, in spite of the complexity of online interactions, most existing research focuses on linking mental health to the amount of engagement online \citep{jaycox2024social}. While this is a useful first step, understanding how online activities shape suicide risk requires focusing not only on time spent online but also on the types of online activities contributing to risk. For example, suicide risk may be related to online activities like consumption of content related to self harm, cyberbullying, and intimate online relationships. While some prior research has examined these behaviors, they were often not studied at scale and linked to mental health or suicidality more generally rather than suicide death \citep{john2018self,yildiz2019suicide,aliverdi2022social}.}

\revision{Using open text summaries of death investigations from a nationwide mortality registry, we use qualitative, statistical, and computational methods to address three aims: First, we conduct a thematic analysis to characterize the types of online activities among 29,124 youth suicide decedents from 2013 to 2022, drawing on the Integrated Motivational-Volitional (IMV) and Durkheim theories of suicide risk \citep{o2018integrated,durkheim2005suicide}. Next, we examine how these online activities vary by decedent characteristics including age, means of death, mental health, and interpersonal problems. Finally, we investigate how these onlines activities varied over time, particularly in the aftermath of the COVID-19 pandemic.} Collectively, our work makes three contributions:

\begin{enumerate}
    \item We develop a set of 12 themes characterizing online behaviors among youth suicide decedents \revision{related to harm to self, harm to others, interpersonal interactions, activity levels online, and life events}.
    \item We create a framework to model these themes with zero-shot learning. Our framework accounts for the fact that many of the themes are conceptually similar (e.g., perpetrator vs. victim of violence) and, therefore, difficult for an LLM to distinguish.
    \item We use the model-identified themes to show, at population scale, that \revision{online activities are associated with factors influencing suicide completion like age, interpersonal problems, and means of death. We also show that online activities among decedents exhibit temporal variation, including many with particularly large increases during the COVID-19 lockdowns.}
\end{enumerate}

\revision{The near-unprecedented pace of near-ubiquitous adoption of online social activities, with the parallel rise in youth suicide in the past 20 years, requires new analytic approaches to bring long-standing theories of suicide risk into the 21st century. While digital platforms have taken some steps to address overt expressions of suicidality (e.g., allowing posts to be flagged by others, linking to crisis lines), these efforts only address indicators of the acute volitional suicidal crisis phase immediately preceding an act of self-harm. In contrast, the themes identified in this work identify indicators of risk more upstream of this crisis phase and therefore offer greater opportunities for intervention. While challenging to implement, our work emphasizes the potential for online platforms to collaborate more systematically with mental health professionals and investigations of suicide deaths in order to identify opportunities for addressing how specific online activities contribute to suicide risk.}

\section{Online Activity and Youth Suicide}

\paragraph{Modeling Suicidality Online.} A large body of literature studies expressions of suicidality in online spaces \citep{macrynikola2021does,mok2015suicide}. An important area of research has been characterizing suicidality on social media, web forums, and other online spaces. A number of papers detect suicidal ideation in social media and other online spaces \citep{skaik2020using}, and use these detectors to perform downstream tasks like modeling transitions from other topics to suicidal ideation \citep{de2016discovering}, studying what types of discourse promotes or deters ideation \citep{de2017language,chancellor2021suicide}, analyzing how suicidality spreads \citep{kumar2015detecting}, and even designing evaluations and systems for online intervention \citep{kavuluru2016classification,sawhney2021towards,mccarthy2010internet}. Other work has used online activity to predict population-scale suicide mortality rates \citep{patel2024predicting,choi2020development}. We build on this important body of work to understand decedents' other online behaviors, apart from expressions of suicidality. 

\paragraph{Suicidality vs. Mortality.}
Importantly, our study examines social media use in the context of suicide \textit{deaths} rather than \textit{suicidality} more generally (i.e., expressions of suicidal ideation, intention, or plans).
While existing social media datasets can be used to track suicidality, these analyses are often limited in their ability to identify suicide death which an epidemiologically distinct phenomenon. First, suicide deaths only occur in a very small fraction of people who experience suicidality. Suicidal thoughts such as feeling persistent sadness or hopelessness are relatively common -- as of 2023, 39.7\% of youth reported these feelings in the prior 12 months \citep{shain2018youth}. A smaller number of youth seriously contemplate suicidal action (20.4\%), and even fewer attempt suicide (9.4\%) with a first attempt-to-completion ratio of 41:1. Second, the population expressing suicidal ideation is distinct from those who die by suicide. For instance, rates of suicide completion significantly differ by sex (male youth have a ratio of 16:1 attempt:completion for the first attempt) and means of death such as firearms drastically increases lethality of a suicide attempt \citep{shain2018youth}. Therefore, while identifying and mitigating suicidality online is an important step, separate research is required to understand whether these findings generalize to prevention of suicide mortality. Our work represents an important first step in this direction; though we do not have the data to estimate suicide risk following different online behaviors, we are able to characterize what online behaviors are linked to suicide death and analyze variation in these behaviors.

\section{Data}

In order to analyze online activities among youth who have died by suicide, we use a registry of suicide and other violent deaths in the U.S.A. Managed by the Centers for Disease Control and Prevention (CDC), the National Violent Death Reporting System (NVDRS) is a comprehensive database of violent deaths (including suicides) in states reporting into the system. The number of reporting states grew over time. In 2013, 17 states reported suicide data to NVDRS, growing to 30 states in 2016, and by 2020 there were 47 reporting states (including DC, excluding CA, FL, HI, and NY). NVDRS contains records of \textit{all} suicide deaths in each reporting state, making it a population-scale dataset. 

Data reported to NVDRS are derived from the death certificate and death investigation reports. Data include decedent's demographics (age, sex, race, etc.), key characteristics of the death (date, location, weapon, etc.), a set of binary-coded contributing circumstances based on the investigation reports (mental health problems, school problems, interpersonal problems, etc.), and two narrative summaries from the perspective of law enforcement \revision{(LE)} and the corner/medical examiner \revision{(CME)}. The narratives are composed by trained abstractors in each state, who summarize the investigation reports to provide detailed descriptions of the contributing circumstances of the decedent's death. These narratives are the primary source of information about online activities, as none of the coded circumstance variables systematically characterize what a decedent does online (the only relevant variable is ``Recent Disclosed Suicidal Thought/Plan via Electronic Means''). NVDRS narratives are based on LE and CME investigations into each death, so information about online activities likely comes from two primary sources: accounts by the decedent's next of kin (family, friends, school, etc.) and searches of the decedent's electronic devices. The official investigation reports detail important findings, and the NVDRS abstractors write a summary for researchers with the goal ``to provide the context for understanding the incident'' \citep{nazarov2019research}. 

We note three key points about how to interpret findings from the narratives: first, the narratives describe a high precision but not high recall set of circumstances. Due to gaps or omissions by next of kin, LE/CME, and abstractors, NVDRS narratives do not provide a comprehensive account of the circumstances salient to each suicide; however, per our conversations with NVDRS abstractors, circumstances referenced in the narratives are likely relevant to the death. Second, the narratives collectively describe a range of circumstances experienced by suicide decedents. NVDRS cannot be used on its own to calculate the suicide risk associated with each behavior, as it provides the prevalence of these behaviors only among decedents and not among the full population. However, NVDRS includes all decedents in reporting states and, therefore, allows us to characterize behaviors among decedents at population scale. Third, since NVDRS relies on death investigations, there is significant heterogeneity in the quality of reporting by decedent characteristics. For instance, decedents who were male and part of some racial/ethnic minority groups tend to have shorter narratives, which often correlates to fewer circumstances reported \citep{mezuk2021not}. Notably, however, younger decedents tend to have longer narratives than older age groups.

Our analysis covers a ten year period, from 2013 through 2022. Our analytic sample consists of all 29,124 single suicide deaths among youth ages 10-24 for whom the circumstances surrounding the death are known in NVDRS. We chose to study decedents ages 10-24 in order to remain consistent with the CDC's definitions of youth \citep{nchs2023suicide}. On average, narratives contain 1509 characters and 19 sentences. Access to the NVDRS Restricted Access Database, containing the narrative data, was approved by the CDC. Our institution's Institutional Review Board deemed this study exempt from human subjects regulations because the NVDRS data is limited to deceased individuals. 

\section{Thematic Analysis of Online Activities}

We start by using the NVDRS narratives to identify a set of themes describing the online activities of youth who died by suicide. \revision{We adopt a mixed analysis approach, combining inductive and deductive coding to develop themes.}

\paragraph{Theory} In the present study, we draw on two key theories to characterize heterogeneity of online activity in the context of suicide risk: IMV theory and Durkheim's theory of suicide. IMV theory explains how certain behaviors may lead someone from triggering events to suicidal thoughts to planning and death \citep{o2018integrated}. IMV theory charts this progression using three risk phases: pre-motivational (precipitating events, stressors, and triggers), motivational (suicidal ideation and intention), and volitional (enaction and completion of suicide). IMV theory also outlines various mechanisms underlying an individual's transition between these states. For instance, decedents may experience \textit{threats to self moderators}, which trap them in the negative emotional states following a stressful event (e.g., withdrawal or intensification of use of an online platform). These negative emotional states may evolve into suicidal ideation if a decedent experiences \textit{motivational moderators} that lessen the likelihood that a decedent sees an alternative to suicide (e.g., online spaces can exacerbate feelings of isolation, evident in instances of sharing private or intimate content). Finally, the entry into the volitional phase is facilitated by \textit{volitional moderators}, or behaviors that normalize death or increase access to injury means (e.g., discussing suicide online). Importantly, while expressing suicidal ideation or intention online is an important risk factor, IMV theory suggests that other risk factors that are observable in online spaces and could also serve as important points of intervention. 

Additionally, Durkheim’s theory of suicide suggests that suicide risk factors are likely to violate established social norms in one of two ways \citep{durkheim2005suicide}. Durkheim suggests social groups function best when members balance their levels of social integration (connection with others) and social regulation (adherence to conventions in communication and behavior). An imbalance -- whether excess or lack of regulation or integration -- can increase the risk of suicide among group members. Unlike IMV theory, which points to potential mechanisms linking a behavior to suicide risk, Durkheim's theory offers a clear heuristic for whether a behavior is a suicide risk factor, which is useful for theme development. Together, these theories allow for the evaluation of \revision{online} experiences which may inform suicide behaviors.

\paragraph{Annotation Sample.} Many NVDRS narratives did not reference online spaces at all and, therefore, were irrelevant to our analysis. To obtain a \revision{high-recall} sub-sample with high density of relevant cases, \revision{we up-sampled narratives for annotation} that contained keyphrases that are likely to reference the decedent’s online behavior. To identify these keyphrases, we started with a starter set of 15 hand-collected phrases \revision{that the team of authors developed based on their knowledge and based on reading a small number of narratives where the ``Disclosed Suicidal Thought/Plan via Electronic Means’’ variable was positive} (e.g., social media, computer, phone). \revision{To ensure we were including other phrases that are used in NVDRS}, we augmented the starter set with the words and phrases that had highest cosine similarity to the starter phrases (e.g., web, devices). \revision{To calculate cosine similarity, we trained word2vec embeddings on all 29,124 narratives in the analytic sample; narratives were split into words and lemmatized using spaCy's tokenizer, and stop words were removed.} Using these embeddings, we identified the 50 most similar words to each starter phrase, and manually filtered low-precision or irrelevant phrases from this list \revision{by examining a small sample of cases containing each phrase} (e.g., ``communicat*'' which referred to both digital and face-to-face communication). 

\revision{This procedure gave us a set of 51 keyphrases relevant to online activities.} 34.4\% of \revision{all NVDRS narratives in the analytic sample} contained one or more of our 51 keyphrases; \revision{the phrases from this set that were most commonly found in the analytic sample} included ``text'' (found in 19.1\% of cases), ``messag*'' (15.1\%), ``post'' (6.4\%), ``social media'' (5.6\%), ``gam*'' (3.4\%), and ``computer*'' (2.8\%). \revision{We annotated 645 cases in total, including 545 cases containing one or more of the 51 keyphrases and 100 cases that did not contain any keyphrases. We estimate the recall of the keyphrases in identifying cases where online activities are mentioned is above 0.973 and the precision is 0.804 (Appendix: Evaluation).} 

\paragraph{Methods.} We performed a thematic analysis of the online activities described in NVDRS narratives of youth decedents (aged 10-24), \revision{using a subset of 545 cases containing one or more of the keyphrases and 100 cases containing none of the keyphrases.} Annotators were a group of students, researchers, and \revision{faculty} at a large public USA university. Annotators are all members of a cross-disciplinary epidemiology research group and are affiliated with a range of departments including public health, data science, information, firearm injury prevention, and social work. \revision{Following the mixed-analysis approach of \citet{bingham2023data}, themes were identified through multiple rounds of inductive and deductive coding.} The first \revision{deductive} step of the thematic analysis was attribute coding to determine whether each narrative referenced any of the decedent's online activities. Online activities include internet platforms (e.g., social media, search engines, forums), computer-mediated communication (e.g., chat, texting, video calls, email), or other activities that are likely to require the internet (e.g., video games, pornography). It excluded the use of likely offline communication technologies (e.g., phone calls, voicemail) or electronic devices (e.g., photos or documents stored on a phone/laptop); it also excluded online activities that the decedent was not actively involved in (e.g., someone else using the decedent's phone to track their location postmortem). 72.6\% of cases in the annotation subset \revision{(468 of 645 annotated cases)} referenced the decedent's online activities in a way that met our inclusion/exclusion criteria. 

After identifying potentially relevant cases, the second step required each annotator to conduct an initial \revision{inductive}, exploratory review of 20 narratives and open-coding of any engagements in online spaces (e.g., ``posting about depression,'' ``consuming sexually explicit content''). Then, the annotators collaboratively developed code definitions by discussing emerging ideas in the data. \revision{In additional inductive rounds, annotators coded an additional 380 cases (40-50 each), where} they continued exploratory open-coding and started grouping these behaviors into broader \textit{themes} that emerged from the data (e.g., ``personal sharing,'' ``consuming content involving harm to others''). 

\revision{The third step was deductive, with the same set of 400 unique narratives as step 2.} Annotators applied IMV and Durkheim’s theories to refine the themes. Themes were clustered relative to their relationship to one or more risk factors in the IMV theory and violation of norms of either integration or regulation from Durkheim's theory. For example, activities related disclosure and graphic disclosure of suicidality are within distinct themes because graphic disclosure offers imagery of the death which is a different type of volitional moderator; however, disclosures of plans and intents are not separated into distinct themes because they are both volitional moderators by stating intentions. Our team identified several types of online activities discussed in the narratives that did not fit within IMV or Durkheim's theory -- for instance if a narrative referenced normative online behaviors unrelated to any of the IMV risk categories (e.g., using computer-mediated communication to speak with friends or family, playing video games online, listening to music, online dating, or attending remote school or therapy appointments); these activities are not included in any of the themes. 

The team of annotators conducted five rounds of iterative inductive and deductive coding, cycling between steps 2 and 3, before finalizing themes. In a final verification step, annotators labeled a new set of \revision{245 unique} narratives with the codebook and discussed any uncertain instances. \revision{In total, the thematic analysis was conducted with 645 narratives.}

\paragraph{Results.} Our analysis reveals 12 themes describing decedents' online activities, as depicted in Figure~\ref{fig:teaser}. \revision{As detailed below, each theme is related to a different phase of risk from the IMV model (shown in yellow, orange, and red in the top row of boxes in Figure~\ref{fig:teaser}) and a different transgression of Durkheim's norms (shown in light green and light blue in the bottom row of boxes).} A detailed description of each theme is given in Table~\ref{tab:themes}.

We identified three themes related to \textbf{Harm to Self} \revision{(right-most purple box in Figure~\ref{fig:teaser})}: \textit{graphic disclosure} (disclosing suicidal intent by posting photos/videos of the suicide in progress or of the means of death), \textit{disclosure} (non-graphic \revision{ways of disclosing suicidality online, including direct references or} allusions to ideation or intent), and consuming \textit{content about self-harm and suicide} (including web searches about means of death, participation on suicide forums, watching suicidal videos, etc.). These behaviors transgress norms of regulation (e.g., most sites have policies against discussions of harm or violence), and are likely linked to volitional moderators that help decedents identify their means of death and reduce fears about their suicide.

Next, we identified three themes related to \textbf{Harm to Others} \revision{(second purple box from the right in Figure~\ref{fig:teaser})}: where the decedent is the \textit{perpetrator} of harm (e.g., decedent bullies someone, engages in illicit activities, or creates/posts violent or explicit content unrelated to their own suicide); where the decedent is the \textit{victim} of harm; and where the decedent consumes \textit{content showing harm to a real or fictional other} (including violent video games, watching videos of violence, etc.). These behaviors also transgress norms of regulation as they violate policies of many online sites, and are likely linked to volitional moderators that may desensitize someone to violent imagery and thoughts of death. 

We found another set of three themes related to \textbf{Interpersonal} interactions \revision{(middle purple box in Figure~\ref{fig:teaser})}: including \textit{online conflicts} (both conflicts that initiate/progress online and conflicts about online activities); \textit{personal sharing} (sharing of private or intimate content, negative emotional states, etc.); and \textit{online relationships} (i.e., interactions with an individual who they primarily know online for long duration or with a high degree of personal sharing, including romantic relationships, friendships, sex solicitation, etc.). These online behaviors may represent threats to self moderators and transgress norms of integration, since they may involve greater intimacy than is typical in online spaces.

Two themes related to \textbf{Activity Levels} online \revision{(second purple box from the left in Figure~\ref{fig:teaser})}: \textit{withdrawal} from, cessation of, or low levels of online activities (including cases where a device is taken away because of a disciplinary act from a guardian); and, in contrast, \textit{intensifying} or high levels of online activities. These behaviors transgress norms of integration by increasing or decreasing the frequency of online interactions. They may be linked to motivational moderators that convey underlying themes of thwarted belongingness or to threats to self moderators as they indicate a drop in social problem solving skills.

One final theme \revision{related to \textbf{Life Events and Stressors}} emerged from the thematic analysis \revision{(left-most purple box in Figure~\ref{fig:teaser})}: \textit{problems with online schooling}. \revision{These problems often violate norms of integration and regulation, as students had less interaction with peers and had to adjust to the new norms of online education \citep{edUSEducation}. 
} 

The broad range of behaviors captured by our themes suggests that there are several key points of intervention that are often overlooked in analyses of suicide risk online. This includes a set of activities that can increase the risk of suicide at each step from triggering events to suicide death. While some of these behaviors have already been studied in the literature, studying online trace data in isolation often fails to capture the context of online behaviors that leads to suicide \citep{jaycox2024social}. Using NVDRS narratives allowed us to more directly examine each behavior relative to a confirmed suicide and to examine the trends within confirmed suicides rather than suicidality more generally. 

\section{Modeling Themes Related to Online Activities}

After developing these 12 themes, we next used zero-shot learning to estimate their frequency in the full sample. 


\begin{table*}[t]
\centering

\begin{tabular}{l|l|c|c|c|c}
\textbf{Category} & 
\textbf{Theme} & 
\textbf{IAA ($\alpha$)} & 
\textbf{Performance (P/R/F)} & 
\textbf{Fraction} & 
\textbf{Fraction Adj.} \\ \hline \hline
\textbf{Harm to Self} & Graphic Disclosure & 0.67 & 76.9 / 53.6 / 63.1 & 0.0637 & 0.0912 \\
 & Disclosure & 0.65 & 90.4 / 75.9 / 82.5 & 0.4369 & 0.5203 \\
 & Self-Harm Content & 0.64 & 65.6 / 91.7 / 76.4 & 0.1128 & 0.0807 \\ \hline
\textbf{Harm to Others} &
Perpetrator & 0.52 & 61.9 / 61.9 / 61.9 & 0.0492 & 0.0492 \\
 & Victim & 0.80 & 72.0 / 62.1 / 66.7 & 0.0355 & 0.0411 \\
 & Other-Harm Content & 0.77 & 100 / 58.3 / 73.7 & 0.0119 & 0.0204 \\ \hline
\textbf{Interpersonal} & 
Online Conflict & 0.81 & 67.2 / 74.5 / 70.7 & 0.1257 & 0.1134 \\
 & Personal Sharing & 0.41 & 41.2 / 28.0 / 33.3 & 0.0357 & 0.0525 \\
 & Online Relationships & 0.75 & 68.8 / 68.8 / 68.8 & 0.0032 & 0.0032 \\ \hline
\textbf{Activity Levels} &
 Withdrawing or Low Use & 0.60 & 91.7 / 75.0 / 82.5 & 0.0475 & 0.0581 \\
 & Intensifying or High Use & 0.80 & 88.2 / 51.3 / 65.6 & 0.0173 & 0.0297 \\ \hline
\textbf{Life Events} &
 Problems with Online School & 0.57 & 68.4 / 81.3 / 74.3 & 0.0240 & 0.0202 \\
\end{tabular}
\caption{Performance in detecting themes describing online activities by youth who died by suicide. For each theme, we report the interannotator agreement among the 8-person annotation team (IAA), the performance of the zero-shot classification pipeline \revision{(Performance: P = precision, R = recall, F = F1 score)}, and the fraction of youth decedents the LLM identified containing the theme (Fraction). \revision{Since some themes have imbalanced precision and recall, we also report an estimate of the true fraction, adjusted for precision and recall (Fraction Adj).}}
\label{tab:performance}
\end{table*}

\paragraph{Methods.} Per the terms of the data use agreement, the NVDRS narratives can only analyzed with open-source language models served on a pre-approved subset of our institution's computing infrastructure. Our experiments test instruction-finetuned versions of four of the highest-performing open-source LLMs to date: Llama3.1-8B-Instruct \citep{dubey2024llama3herdmodels}, Qwen2.5 \citep{qwen2}, Phi 3 \citep{abdin2024phi3technicalreporthighly}, and Mistral-7B-Instruct-v0.3 \citep{jiang2023mistral7b}. Given computing constraints, we opt to use the smallest versions of each of these models. Since these smaller LLMs are the most accessible to the public health research community, an important contribution of this work is developing a prompting framework that to effectively identify complex themes using these LLMs. Models are run in a CUDA 12.4 environment, using Python 3.11.9 with one NVIDIA RTX A6000 GPU. 

\paragraph{Pipeline.} 

We developed an iterative prompting approach to identify and classify decedent's online engagement in the narratives. The pipeline is described below. The full text of all prompts is given in Appendix: Prompts.

\textit{Step 1: Splitting Narratives into Sentences.} First, we prompted the LLM to break the narratives into individual sentences. This accounts for the LLM's difficulty in processing the full narrative due to their smaller context length (narrative are often 20+ sentences). For instance, when prompted using a full narrative, the model had much lower recall in recognizing online activities in Step 2 of the pipeline (52.4 full narratives vs 94.5 sentences).

\textit{Step 2: Identifying Online Activities.} 
Second, we prompted the LLM to identify references \revision{to} online spaces using a four-step chain of thought prompt. This accounted for the LLM's poor performance in distinguishing between online behaviors of interest identified in our thematic analysis and their offline counterparts (e.g., including an offline argument in the online Conflict theme). Without this step in the pipeline, the precision of most themes dropped significantly (e.g., Online School dropped from 68.4 to 55.2).

\textit{Step 3: Identifying Sets of Related Sentences.}
Third, we prompted the LLM to identify whether pairs of sentences were gramatically or conceptually related to each other. While most sentences can be correctly interpreted on their own (e.g., ``V was last seen alive in the late evening the day prior. V's grandmother woke up in the early morning and went to check on him. V was found on his bed with an apparent self-inflicted [means of death].''), this accounted for the fact that sometimes the interpretation of one sentence changes in light of the sentence before or after it. Without this step in the pipeline, the recall of a few themes dropped slightly (e.g., Victim).

\textit{Step 4: Delineating Themes.}
Finally, we introduced a novel prompting strategy, inspired by decomposition prompting, for delineating similar themes \citep{schulhoff2024prompt}. We crafted multiple-choice prompts for a set of 3-6 themes, to create clear inclusion and exclusion criteria for each code. For instance, the LLM was often unable to distinguish between various aspects of online engagement in these narratives (e.g., someone posting ``goodbye'' on Facebook was classified as Withdrawal instead of Disclosure). The multiple choice prompt offered the LLM various options for how to classify a sentence and forced the LLM to choose between options it may otherwise view as similar. Notably, the model's performance did not increase when we allowed the model to select multiple options from the list; a single select prompt was required for the model to have this forced choice and properly apply the inclusion/exclusion criteria for each code. We crafted four prompts: one whose options corresponded to the Harm to Self and Harm to Others categories (i.e., depending on which option the LLM selected, the sentence would be classified into one of the six themes under these categories), one for the Interpersonal category;  one for the Activity Level and Life Events themes; \revision{and one to identify the source of information used to get information about online activities}. Three authors manually tuned these prompts by refining the options in the multiple choice question using 30\% of the annotated data; performance was tested on the remaining 70\% of the data. Even themes like Online School that seem distinct from the remaining themes benefited from this multiple choice prompting approach (F1 score of 48.4 with a similarly tuned chain of thought prompt vs. 74.3 in multiple choice). 

\paragraph{Evaluation.} \revision{We evaluated the performance of our pipeline using the 645 unique narratives manually annotated in the thematic analysis}, including a subset of 105 multiply-annotated narratives by 2-4 annotators for inter-annotator agreement (760 annotations in total). To evaluate the performance of the full pipeline, annotators assigned themes after reading the entire narrative. However, annotators also evaluated Steps 1-3 of the model by annotating these characteristics one sentence at a time.

\paragraph{Results.} Steps 1-3 of the pipeline achieve high performance at their respective tasks. In Step 1, we verify that the narrative is split into valid sentences (we reviewed over 18,000 sentences during annotation and all were valid), and that the narrative is split into sentences without further alteration (stringing the edit distance on average less than 1\% of the length of the narrative). In Step 2, the model closely matches human annotations in detecting whether the narrative mentions any online activities (precision 95.8, recall 94.5, F1 95.1). In Step 3, the model has high recall in pairing related sentences (88.3), though it also has relatively poor precision (53.7).

The LLM identified 9,194 of 29,124 (32.9\%) cases in the analytic sample where any of the decedent's online activities are described. Since over 97\% of American youth report using the internet daily \citep{pewresearchTeensSocial}, this result suggests online activities may be under-reported in NVDRS narratives. Using the prompt from Step 2, we can detect what types of technologies are referenced by the narrative. Of the cases that describe online activities, the majority referenced messaging (text, chat, etc.) and devices (phones, laptops, etc.), with about 15\% referencing social media or gaming. There is variation in how often different types of technology are referenced for each theme (Figure~\ref{fig:types}).

\revision{In addition to the type of technology, over 60\% of the time narratives explicitly report the sources used to get information about a decedent's online activities. We find that 10.7\% of cases obtain their information from searches of the decedent’s technology, 54.8\% explicitly stated that they get information from reports from next of kin, 34.5\% implicitly suggest that they may get information from next of kin (e.g., by referencing their communication with the decedent), and 10.6\% have an unspecified source of information. Our prompt achieved 86\% accuracy in modeling information sources, with most errors conflating implicit and explicit references to next of kin (Table~\ref{tab:source-performance}). Our results suggest that death investigations are often not systematically obtaining information about decedent’s online activities through searches of their technology. Instead, most are relying on reporting from next of kin, who may have more context for the decedent's actions but an incomplete picture of the decedent's online life.}

Table~\ref{tab:performance} shows the performance of the entire pipeline by theme. Our main results were calculated using Llama3.1-8B-Instruct, as the other LLMs had worse performance (not shown). Most themes have F1 scores \revision{between 65 and 75. The moderate performance of our prompts is likely due to the inherent ambiguity in applying complex constructs to diverse narratives; the difficulty of the task is evidenced by our consistently moderate inter-annotator agreement scores among the authors.} Accordingly, the two themes with poor F1 scores (Personal Sharing and Perpetrator) also have poor inter-annotator agreement. Since some themes have imbalanced precision and recall \revision{(typically higher precision than recall, an overall undercount)}, we report the raw fraction of cases containing each theme from the LLM predictions as well as an estimate of the true fraction, adjusted for precision and recall. \revision{Let $F$ be the fraction of cases the LLM identifies as being related to a given theme and let $P$ be the precision and $R$ be the recall of the LLM in identifying this particular theme in the full dataset. The fraction adjusted for precision and recall is $F \cdot P / R$, which approximates the true fraction of the cases related to the theme, including the LLM’s false negatives and excluding the LLM’s false positives.}

\section{Temporal Variation}

\begin{figure}
    \centering
    \includegraphics[width=\linewidth]{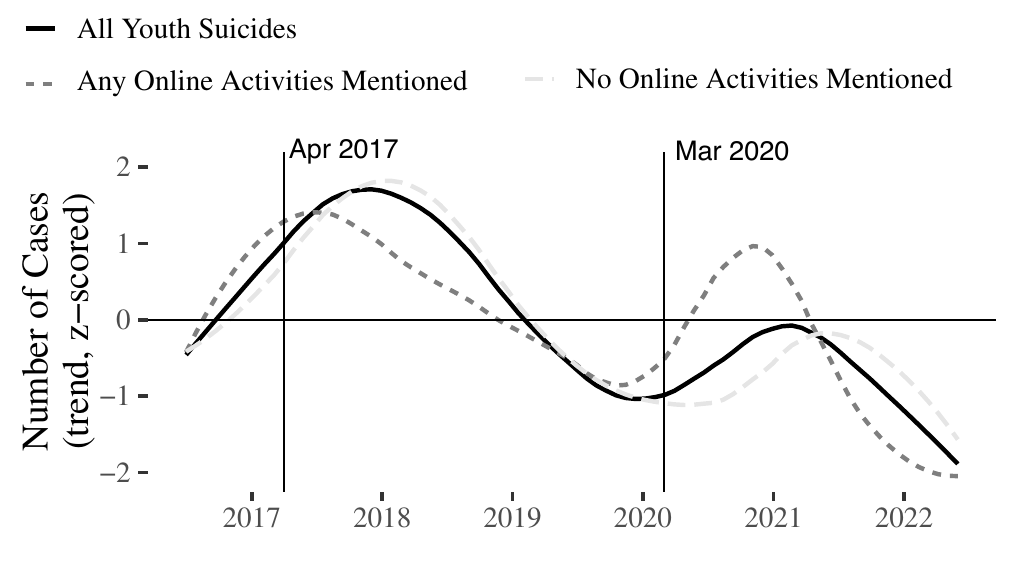}
    \caption{\revision{There are distinct temporal trends in the number of suicides among youth (black solid line), the number of narratives describing online activities of youth suicide decedents (dark gray dotted line), and the number of narratives that do not describe online activities of youth suicide decedent (light gray dashed line). For ease of comparison, we plot the z-scored trend of each time series.}}
    \label{fig:ts-suicides}
\end{figure}

\begin{figure*}
    \centering
     \begin{subfigure}{0.45\textwidth}
         \centering
         \includegraphics[width=\textwidth]{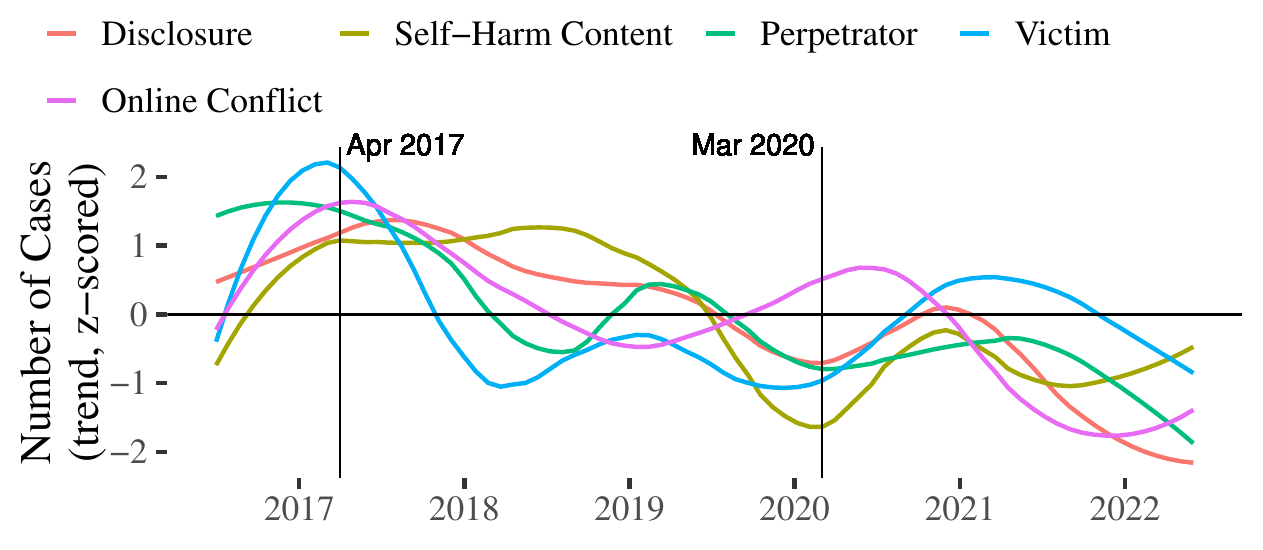}
         \caption{Themes w/ peak around 2017}
     \end{subfigure}
     \begin{subfigure}{0.45\textwidth}
         \centering
         \includegraphics[width=\textwidth]{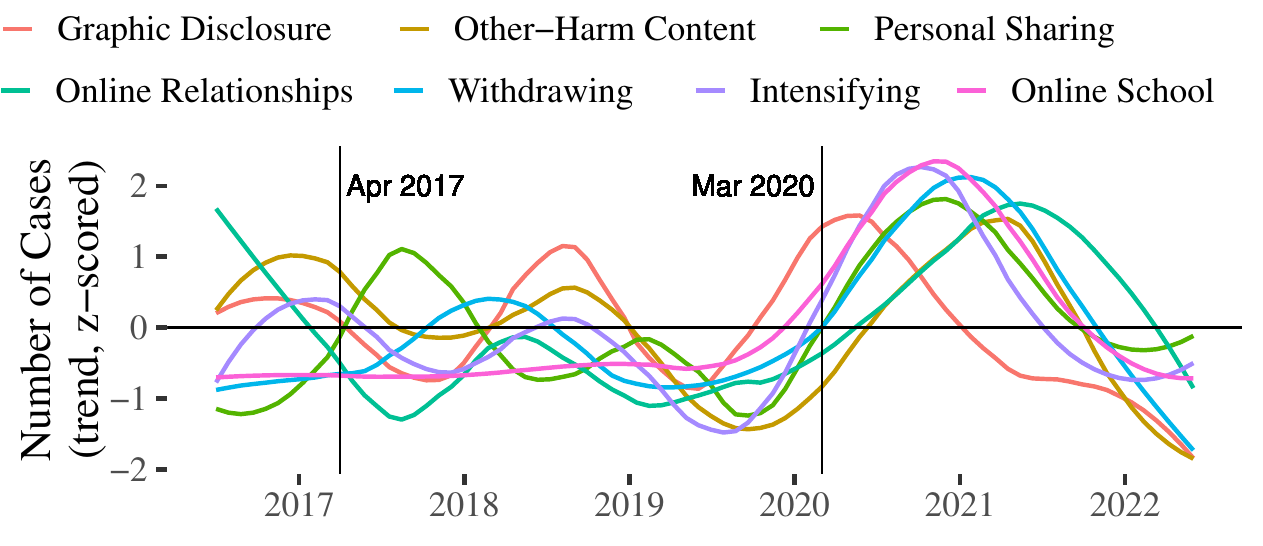}
         \caption{Themes w/ peak around 2020}
     \end{subfigure}
    \caption{\revision{Two sets of temporal trends among the 12 themes describing online activities of youth suicide decedents: (a) themes with peaks around April 2017 and (b) themes with peaks immediately after March 2020.} Each time series represents the \revision{number} of decedents with the theme. We plot the z-scored trend of each time series. \revision{A plot for each individual theme is given in Figure~\ref{fig:ts-all}.}}
    \label{fig:ts}
\end{figure*}

Using our fully labeled dataset, we analyze the variation in references to online activities and in each theme over time. Current literature has noted seasonality in youth suicides, with the highest rates of suicide deaths in the spring and fall months \citep{hatton2024temporal}. Temporal trends of factors that contextualize adolescent suicide are equally important to examine because they can reveal patterns of increasing warning behaviors, which can inform prevention programs aimed at raising awareness of these behaviors as potential warning signs. This is particularly salient for experiences which occur through social media, given the rapid pace at which social media tools and normative interactions change within social media.

\paragraph{Methods} First, we analyze the temporal trend in how often NVDRS narratives of youth decedents reference \textit{any} of their online activities (i.e., the cases identified in Step 2 of the pipeline, not those from just one specific theme). To do this, we construct a monthly time series of the \revision{number} of decedents with narratives referencing online spaces \revision{and, for comparison, the number of decedents with narratives that don't reference online spaces}. For consistency, the time series includes only decedents from states that reported into NVDRS during the entire period; we start the temporal analysis in 2016 instead of 2013, since there were 30 states reporting into NVDRS in 2016 instead of just 17 in 2013. Using the method of moving averages, each monthly time series is decomposed into trend, seasonal, and random components to create a clearer picture of the temporal variation in the trend without the added noise created by seasonality. \revision{To facilitate standardized comparison of time series with different scales, each trend is z-scored.} We also analyze the temporal trend in each of the individual themes using the same approach. 

\paragraph{Results} \revision{Figure~\ref{fig:ts-suicides} shows the z-scored trends in the number of youth suicide in the 30 states reporting into NVDRS from 2016 to 2022, relative to whether the narrative mentioned online activities. Overall, there was a large spike in the trend of youth suicides in early 2018 and a second but slightly smaller increase in 2021. However, the trend in youth suicides related to online activities exhibited a different pattern, with two prominent spikes: one in early 2017 and a second prominent one in late 2020. Although it is not possible to causally link any external factors to the trends, both of these spikes are concurrent with major national events. The first spike corresponds to a period of time when several major sociopolitical events occurred, including the \#MeToo movement, the Women's March, and the start of President Donald Trump's first term. It also coincided with the international release of TikTok in late 2017. These events corresponded to greater social engagement online. Around this time, the Netflix series \textit{13 Reasons Why} was released, which has been linked to suicide deaths among youth \citep{bridge2020association}. This first spike is present among all youth, suggesting the sociopolitical climate may have affected youth offline and online. The second spike corresponds to the lockdowns and school closures during the COVID-19 pandemic \citep{edUSEducation}. This period marked a major structural shift in how youth engaged in online spaces, including increased screen time due to the transition to remote schooling, connecting with friends, and screen-based entertainment. It was also associated with increases in mass online sharing of urgent, unverified health information that likely increased anxiety and uncertainty. This second spike was significantly more prominent among decedents with online activities, likely because of the significant shift online during this period.}

Figure~\ref{fig:ts} shows the trends for all themes. For ease of visual inspection, trends are heuristically grouped into two plots by the timing of their main peak. \revision{A number of themes contribute to the 2017 peak in youth suicides and had a general decline in frequency from 2017 to 2022, including disclosing suicidality online, perpetration of harm, victimization, and online conflicts (Figure~\ref{fig:ts}a).} This finding is somewhat surprising, because online disclosures and cyberbullying remained common during this period \citep{livingstone2016cyberbullying,pourmand2019social} and because 
the amount of time spent online by adolescents also increased between 2016 and 2022. Since adolescent suicidal ideation and attemps also increased during this time \citep{brener2024overview}, we would anticipate an increase in disclosures and interpersonal issues among decedents over time \citep{pewresearchTeensSocial}. 
One explanation is that the trend is driven by the major dip during 2020, when LE/CME were also less likely to formally review decedents electronic devices (Figure~\ref{fig:ts-source}). This likely occurred in relation to social distancing, when LE/CME would avoid entering homes during lock down. The mirrored trends between LE/CME review and other themes underscores the importance of systematic review for elucidating and accurately accounting for these exposures in death reporting. 
Another explanation for the decline in reported disclosures and interpersonal issues is that youth have more control and ownership over who sees their online profiles, and therefore, it is harder for next of kin and LE/CME to see disclosures in \revision{online spaces}. Additional research on motivations and means of \revision{online} disclosures and interpersonal issues would help clarify the trend in the present study.

Many themes exhibited a sharp increase between 2020 and 2021, including school problems, withdrawing or intensifying time spent in online spaces, and online relationships (Figure~\ref{fig:ts}b). \revision{This trend likely reflects the increased time spent online during the COVID-19 lockdowns, as well as some of the unique pressures that accompanied the shift from in-person to digital spaces.} For instance, the online shift may have increased the frequency and saliency of interpersonal interactions online, including online relationships and withdrawals from online spaces. During a time when there was high amounts of societal uncertainty and anxiety, having troubles with online schooling likely exacerbated existing feelings of distress. Additionally, USA adolescents were found to spend on average 40\% more time on social media for socialization and entertainment \citep{marciano2022digital}. Spending more time online during lockdown has been linked to declining mental health \citep{pantic2014online} and likely exacerbated feelings of distress by exposing individuals to higher amounts of COVID-19 information and misinformation. 
While the lockdowns have ended and are unlike to reoccur in the coming years, lessons can be drawn from these findings in the case of future events that increase the necessity of spending time online.


\section{Heterogeneity by Decedent Characteristics}

\begin{figure}
    \centering         \includegraphics[width=\linewidth]{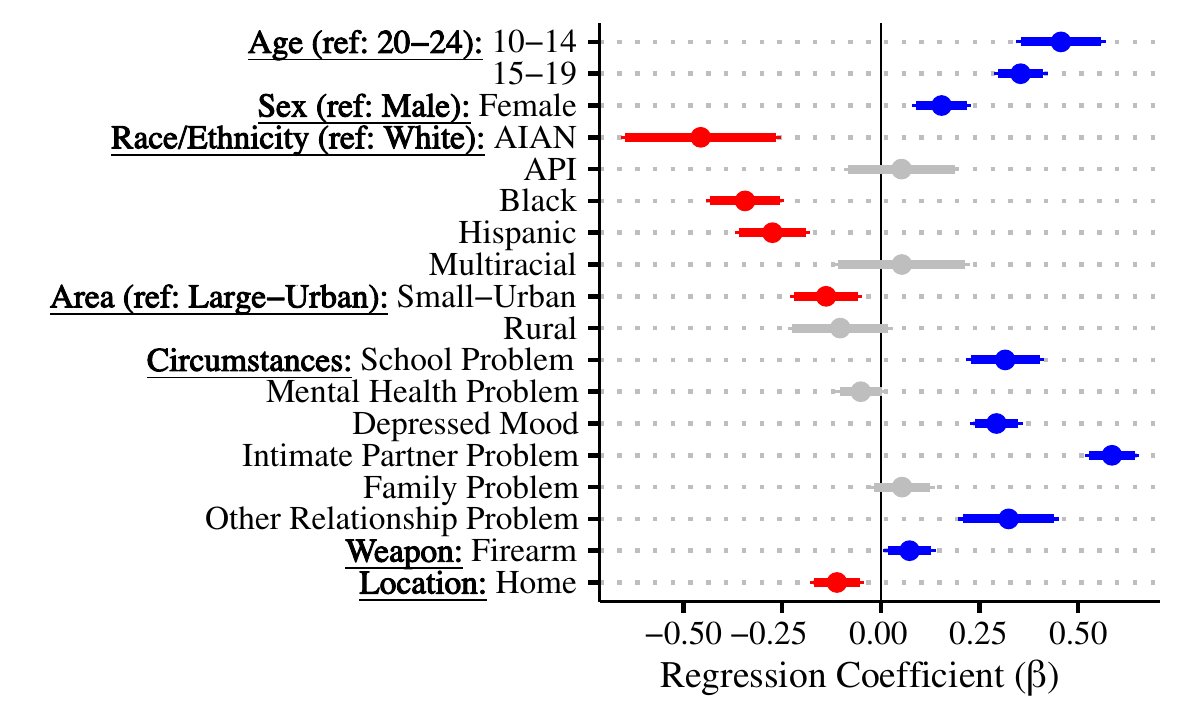}
    \caption{The association between whether the narrative mentioned any online activities (dependent variable) and the decedent's demographics and circumstances (independent variables). This plot visualizes regression coefficients and 95\% confidence intervals. Error bars are adjusted for multiple comparisons using a Bonferroni correction.}
    \label{fig:reg}
\end{figure}

\begin{figure*}
    \centering
    \includegraphics[width=0.85\textwidth]{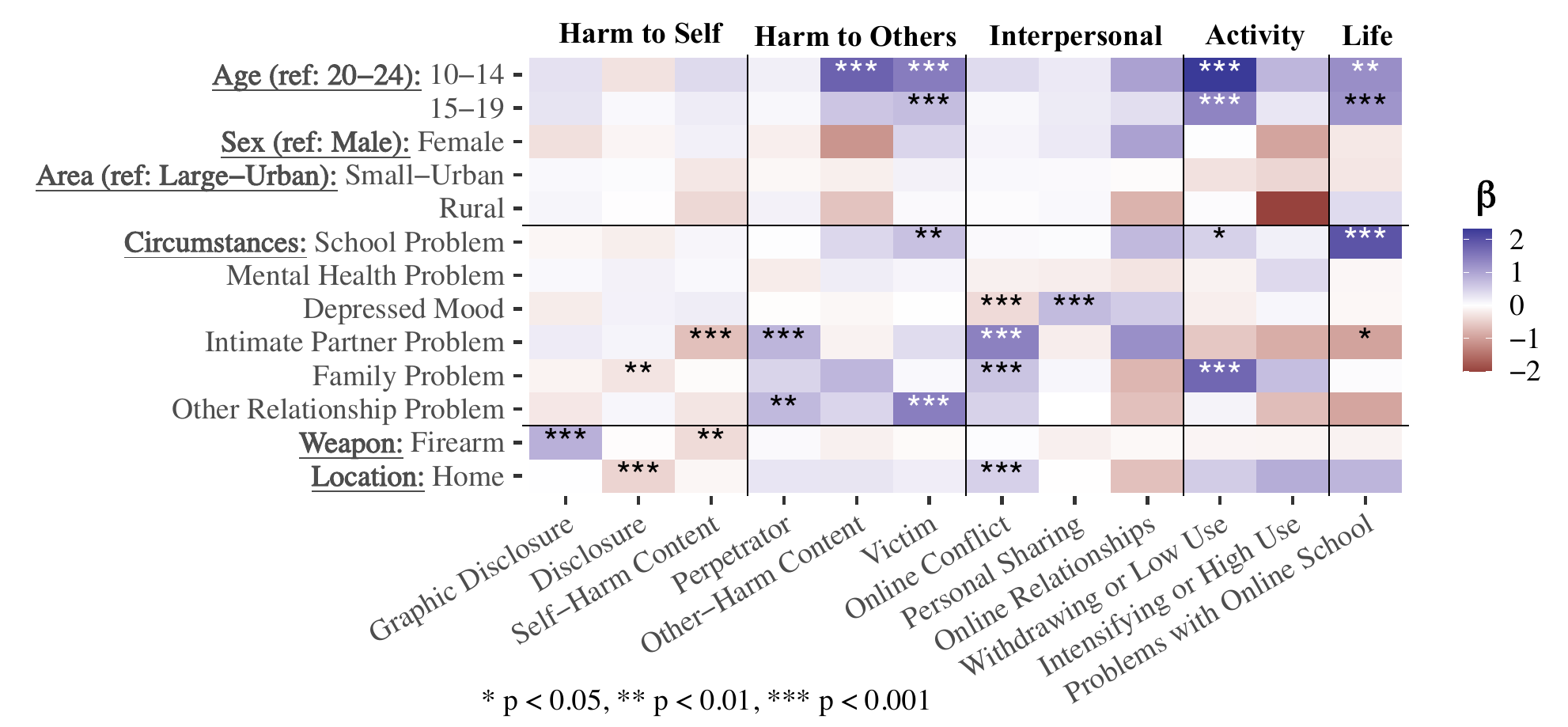}
    \caption{\revision{The association between each theme (dependent variable) and the decedent's demographic characteristics and circumstances (independent variables). For each theme, a logistic regression is used to calculate these associations. The heatmap visualizes the magnitude and direction of the regression coefficients, with darker blue representing larger positive coefficients, darker red representing larger negative coefficients, and white representing coefficients closer to 0. Asterisks represent the significance of each coefficient (* $p<0.05$, ** $p<0.01$, *** $p<0.001$), with p values adjusted for multiple comparisons using a Bonferroni correction. The exact regression coefficients are given in Figure~\ref{fig:reg-table}.}}
    \label{fig:heatmap}
\end{figure*}

Finally, we examine the decedent's demographic and contextual variables to better understand trends related to \revision{online spaces} and youth suicide. 

\paragraph{Methods} We fit a logistic regression model to predict each theme, using several characteristics of the decedents as covariates: demographics (age group, sex: male/female, gender: trans/not trans, race, military status), death characteristics (month, day of week, location: home/not home, weapon: firearm/other), and NVDRS's abstracted binary-coded circumstances related to the suicide (problems with school, mental health, depressed mood, intimate partner, family, and other relationships). We also use fixed effects to control for three factors related to how the narrative was written: the state and year where the data was abstracted, \revision{the source of information used to obtain information about online activities (next of kin, LE searches, or unknown) (Figure~\ref{fig:sources}),} and the narrative length. These characteristics are provided in the NVDRS metadata, and demographic characteristics are often taken from the death certificate. To avoid issues of multicollinearity, we verified that all variance inflation factors are less than 3. Error bars and p values are adjusted for multiple comparisons using a Bonferroni correction

\paragraph{Results} Figure~\ref{fig:reg} shows the results of the regression for all mentions of online activity, while Figure~\ref{fig:heatmap} shows a heatmap summarizing the coefficients across all 12 themes; \revision{the magnitude and direction of each coefficient is visualized in color (larger positive coefficients in darker blue, larger negative coefficients in darker red), with asterisks representing their significance.} For ease of interpretation, we control for but do not show a number of covariates that do not have consistent patterns of significance (e.g., race, transgender, military).

Across all mentions of \revision{online spaces} (Figure~\ref{fig:reg}), the decedent being female, being a student, having a problem at school, having depressed mood, having an intimate partner problem, and using a firearm for means of death were all positively associated with \revision{mentions of online spaces} in the narrative, while the location of death being home; being in the younger age groups of 10-14 and 15-19 (as compared to 20-24) was also positively associated with mentions of \revision{online spaces} in the narrative. Several trends align with extant knowledge and literature. For example, females tend to spend more time in online spaces, and that the association between time spent online and poor mental health was of greater magnitude for females than males \citep{twenge2020gender}. Additionally, younger children tend to experience greater mental health challenges associated with internet use \citep{surgeon2021protecting}.
When examining the specific online experiences, younger decedents were more likely to have viewed content showing harm to others, experienced online forms of victimization (e.g., cyberbullying), withdrawn from online spaces, and had problems with schooling (Figure~\ref{fig:heatmap}). Some of these experiences are also linked to school problems, disproportionately affecting younger decedents who are more likely to still be in school.   
Given the dearth of research on the context or content of \revision{online} experiences, findings of age and gender differences in types of \revision{online} experiences can be useful for public health programming. It also shows how challenges that occur offline (e.g., in school) may spill over into the online domain. However, this dataset only represents decedents, and therefore we cannot conclusively say that these experiences are definite risk factors for suicide.

Adding to the literature on the association between forms of social media exposures and worse mental health \citep{zsila2023pros} are our findings that decedents with narratives that mention \revision{online spaces} were more likely to have had a depressed mood (Figure~\ref{fig:reg}). Interestingly, those with depressed moods were more likely to share negative emotions through \revision{online} spaces (Figure~\ref{fig:heatmap}). Greater public awareness to recognize signs of suicidality being shared through online spaces could be an important factor to prevent youth suicide, particularly when posts are done through private profiles or messaging channels. This is particularly relevant given the rise in youth-oriented anonymous reporting systems \citep{gourdet2021state}, and the tendency for youth to use these spaces to report risk behaviors such as suicidality or firearm-related events of their peers \citep{thulin2024firearm}. Although personal sharing online can be protective for mental health \citep{bremer2005internet}, the references to \revision{online spaces} in NVDRS narratives were often negative in nature; this likely reflects a reporting bias in death investigations, where negative online interactions are more likely to contribute to death and are more likely to be noted in investigation reports. 

For decedents having relational challenges, such problems with intimate partners, family, or someone else, several \revision{themes} related to interpersonal problems were also more likely to occur (e.g., conflicts starting or intensifying through online interactions, perpetrating harm against another person or being victimized online) (Figure~\ref{fig:heatmap}). Contemporary youth are considered digital natives, where online spaces are a major part of their everyday lives. Therefore, it is not surprising that their interpersonal relationships exist and are informed within online interactions. While the relationship between suicide and cyberbullying has been well-studied \citep{john2018self}, we found evidence that those with interpersonal conflicts were at times acting as perpetrators of harm within online spaces. This may reflect the bi-directional role of cyberbullying harassment and bullying, as over a third of youth who report some form of cyberbullying report being both victim and perpetrator \citep{guo2021cyberbullying}. In some cases, \revision{online spaces} may be mirroring a broader set of interpersonal problems that an individual is experiencing. An important step in understanding the context for suicide mortality could be identifying those who are harming others or experiencing harm online as well as identifying their compounding challenges.

Another finding was the increased likelihood of dying by firearm (as compared to another means of death) among those who graphically disclosed their deaths \revision{online} (Figure~\ref{fig:heatmap}). Graphic disclosures included live streaming or video-calling their own death. Often in narratives, this sharing was non-consensual and resulted in a particularly traumatic way of informing a loved one, friends, or others within one’s social network of one’s death. Given the lethality of firearms as a means of suicide as well as the violence of seeing someone being shot, this type of sharing could further exacerbate the emotional, psychological and social effects of loss survivorship \citep{hanschmidt2016stigma}. As the mechanisms of sharing is often within an adolescent’s \revision{online} networks, the streaming of one’s own death could also increase the likelihood of suicide contagion \citep{yildiz2019suicide}. Identifying both the social media behavior (streaming) and the means of death that is highly probable (firearms) could be important for informing future public health campaigns and ensuring there are supportive systems in place for individuals who witness this type of behavior online. However, there exists a tension in USA around the regulation of social media content, with a movement towards less content regulation as of early 2025. Identifying meaningful, efficient and accurate ways to prevent explicitly harmful content from being directed at others (including through non-consensual sharing, like we found in this behavior) remains a critical, but unmet need when it comes to online safety protections for youth in USA. 


\section{Discussion and Conclusions}

Recent rise in youth suicide highlights the urgent need to understand how online experiences contribute to this public health issue \citep{nchs2023suicide}. Our mixed-methods approach responds to this challenge by developing a set of themes focused on risk factors for suicide mortality in online digital spaces. One of the persistent challenges in understanding the effects of \revision{online spaces} on mental health outcomes is the lack of context of \revision{online} experiences, given the tendency for scientists to measure \revision{online} exposure relative to time spent online rather than the content or interactions in the online space \citep{jaycox2024social}. To fill this gap, we identified 12 specific experiences within \revision{online spaces} that were considered by LE/CME or next of kin to be relevant in contexualizing a given adolescent’s death. We then examined the demographic, social, and contextual variables associated with a given \revision{online} experience to better understand trends related to \revision{online spaces} and adolescent suicide. Our work has three key implications.

First, we urge the field to examine the intersection of \revision{online spaces} with suicide deaths. Understanding expressions of suicidality \revision{online} is important given that it is often a high-risk precursor to suicide-related actions during the volitional phase. However, it is also important to identify multiple points of intervention \revision{online} with different relationships to suicide death \citep{surgeon2021protecting}. Our theory-aligned thematic analysis found themes related to multiple phases of suicide risk, including the pre-motivational and motivational phases that are further upstream from suicide. 

Second, our work suggests that there is scope for further research to estimate how often these behaviors occur in the population, the associated risks, and identify effective points of intervention. While our study did not estimate the risks associated with the themes we identified, future work could explore doing linked studies with social media data, browser history, and suicide registries to calculate this risk. \revision{Future work can also use causal inference to examine the association between each of the 12 themes we identified and spatially varying events. This includes recent state legislation such as anti-LGBTQ+ policies, policies affecting access to mental health care like the Affordable Care Act, and policies regarding school closures and online learning during the COVID-19 pandemic. This also includes exposure to large-scale violence like school shootings, including the location, scale, and media coverage of each event.}

Third, we recommend that existing surveillance systems more consistently track online activities among youth suicide decedents. Although nearly all youth use social media and the internet daily, only a third of cases have online spaces mentioned. Additionally, NVDRS appears to significantly undercount suicide disclosures via electronic means among youth; over 90\% of cases of online suicide disclosure identified by our study were not indicated as such in NVDRS. Furthermore, only 11\% of narratives referencing online activities explicitly noted LE/CME’s review of the decedent’s online activity during the investigation. While some narratives may not have mentioned a review that occurred, most cases suggest no review was conducted (e.g., social media factors were identified by family or friends and shared with LE/CME). 
\revision{This work suggests the importance of creating systematic policies and training for LE/CME investigating youth deaths to include an evaluation of the decedent’s social media and personal devices, and to systematically document these investigations in NVDRS narratives.} 

\revision{Importantly, our results need to be interpreted in light of several limitations related to the data. As previously mentioned, the narratives are a high-precision but potentially low-recall catalog of online activities contributing to decedent suicides. Since most narratives only obtained their information on online activities from next of kin rather than from systematic investigations of the decedent's devices, the risks of omissions are particularly high. The quality of narratives has been shown to vary over time and by decedent characteristics \citep{mezuk2021not}, so our findings may not apply as well to all groups of youth. Additionally, our themes are inherently complex and, therefore, the theme-detection framework achieved moderate performance, often with higher precision than recall. Although we were able to adjust for the imbalance in precision and recall in the overall frequencies of each theme, no corresponding adjustment has been made to the subsequent temporal or regression analyses since we lack sufficient annotated data to estimate how errors vary over time or by decedent characteristics.}

\bibliography{aaai25}

\begin{thebibliography}{49}
\providecommand{\natexlab}[1]{#1}

\bibitem[{Abdin et~al.(2024)Abdin, Jacobs, Awan, Aneja, et~al., and Zhou}]{abdin2024phi3technicalreporthighly}
Abdin, M.; Jacobs, S.~A.; Awan, A.~A.; Aneja, J.; et~al.; and Zhou, X. 2024.
\newblock Phi-3 Technical Report: A Highly Capable Language Model Locally on Your Phone.
\newblock arXiv:2404.14219.

\bibitem[{Aliverdi et~al.(2022)Aliverdi, Farajidana, Tourzani, Salehi, Qorbani, Mohamadi, and Mahmoodi}]{aliverdi2022social}
Aliverdi, F.; Farajidana, H.; Tourzani, Z.~M.; Salehi, L.; Qorbani, M.; Mohamadi, F.; and Mahmoodi, Z. 2022.
\newblock Social networks and internet emotional relationships on mental health and quality of life in students: structural equation modelling.
\newblock \emph{BMC psychiatry}, 22(1): 451.

\bibitem[{Bingham(2023)}]{bingham2023data}
Bingham, A.~J. 2023.
\newblock From data management to actionable findings: A five-phase process of qualitative data analysis.
\newblock \emph{International journal of qualitative methods}, 22: 16094069231183620.

\bibitem[{Bremer(2005)}]{bremer2005internet}
Bremer, J. 2005.
\newblock The internet and children: advantages and disadvantages.
\newblock \emph{Child and Adolescent Psychiatric Clinics}, 14(3): 405--428.

\bibitem[{Brener(2024)}]{brener2024overview}
Brener, N.~D. 2024.
\newblock Overview and methods for the Youth Risk Behavior Surveillance System—United States, 2023.

\bibitem[{Bridge et~al.(2020)Bridge, Greenhouse, Ruch, Stevens, Ackerman, Sheftall, Horowitz, Kelleher, and Campo}]{bridge2020association}
Bridge, J.~A.; Greenhouse, J.~B.; Ruch, D.; Stevens, J.; Ackerman, J.; Sheftall, A.~H.; Horowitz, L.~M.; Kelleher, K.~J.; and Campo, J.~V. 2020.
\newblock Association between the release of Netflix’s 13 Reasons Why and suicide rates in the United States: An interrupted time series analysis.
\newblock \emph{Journal of the American Academy of Child \& Adolescent Psychiatry}, 59(2): 236--243.

\bibitem[{Chancellor et~al.(2021)Chancellor, Sumner, David-Ferdon, Ahmad, and De~Choudhury}]{chancellor2021suicide}
Chancellor, S.; Sumner, S.~A.; David-Ferdon, C.; Ahmad, T.; and De~Choudhury, M. 2021.
\newblock Suicide risk and protective factors in online support forum posts: annotation scheme development and validation study.
\newblock \emph{JMIR mental health}, 8(11).

\bibitem[{Choi et~al.(2020)Choi, Sumner, Holland, Draper, Murphy, Bowen, Zwald, Wang, Law, Taylor et~al.}]{choi2020development}
Choi, D.; Sumner, S.~A.; Holland, K.~M.; Draper, J.; Murphy, S.; Bowen, D.~A.; Zwald, M.; Wang, J.; Law, R.; Taylor, J.; et~al. 2020.
\newblock Development of a machine learning model using multiple, heterogeneous data sources to estimate weekly US suicide fatalities.
\newblock \emph{JAMA network open}, 3(12).

\bibitem[{Curtin and Garnett(2023)}]{nchs2023suicide}
Curtin, S.; and Garnett, M. 2023.
\newblock Suicide and homicide death rates among youth and young adults aged 10–24: United States, 2001–2021. NCHS Data Brief, no 471.
\newblock Technical report, National Center for Health Statistics.

\bibitem[{De~Choudhury and Kiciman(2017)}]{de2017language}
De~Choudhury, M.; and Kiciman, E. 2017.
\newblock The language of social support in social media and its effect on suicidal ideation risk.
\newblock In \emph{Proceedings of the international AAAI conference on web and social media}, volume~11, 32--41.

\bibitem[{De~Choudhury et~al.(2016)De~Choudhury, Kiciman, Dredze, Coppersmith, and Kumar}]{de2016discovering}
De~Choudhury, M.; Kiciman, E.; Dredze, M.; Coppersmith, G.; and Kumar, M. 2016.
\newblock Discovering shifts to suicidal ideation from mental health content in social media.
\newblock In \emph{Proceedings of the 2016 CHI conference on human factors in computing systems}, 2098--2110.

\bibitem[{Dubey et~al.(2024)Dubey, Jauhri, Pandey, Kadian, Al-Dahle, Letman, and Mathur}]{dubey2024llama3herdmodels}
Dubey, A.; Jauhri, A.; Pandey, A.; Kadian, A.; Al-Dahle, A.; Letman, A.; and Mathur, A. 2024.
\newblock The Llama 3 Herd of Models.
\newblock arXiv:2407.21783.

\bibitem[{Durkheim(2005)}]{durkheim2005suicide}
Durkheim, E. 2005.
\newblock \emph{Suicide: A study in sociology}.
\newblock Routledge.

\bibitem[{Faverio and Sidoti(2024)}]{pewresearchTeensSocial}
Faverio, M.; and Sidoti, O. 2024.
\newblock {T}eens, {S}ocial {M}edia and {T}echnology 2024.
\newblock \url{https://www.pewresearch.org/internet/2024/12/12/teens-social-media-and-technology-2024/}.
\newblock [Accessed 16-01-2025].

\bibitem[{Gebru et~al.(2021)Gebru, Morgenstern, Vecchione, Vaughan, Wallach, Iii, and Crawford}]{gebru2021datasheets}
Gebru, T.; Morgenstern, J.; Vecchione, B.; Vaughan, J.~W.; Wallach, H.; Iii, H.~D.; and Crawford, K. 2021.
\newblock Datasheets for datasets.
\newblock \emph{Communications of the ACM}, 64(12): 86--92.

\bibitem[{General(2021)}]{surgeon2021protecting}
General, U.~S. 2021.
\newblock Protecting youth mental health: The U.S. Surgeon General’s advisory on youth mental health.
\newblock Technical report, U.S. Department of Health and Human Services.

\bibitem[{Gourdet et~al.(2021)Gourdet, Kolnik, Banks, and Planty}]{gourdet2021state}
Gourdet, C.; Kolnik, J.; Banks, D.; and Planty, M. 2021.
\newblock State legislation to explore or establish school safety tip lines.

\bibitem[{Guo, Liu, and Wang(2021)}]{guo2021cyberbullying}
Guo, S.; Liu, J.; and Wang, J. 2021.
\newblock Cyberbullying roles among adolescents: a social-ecological theory perspective.
\newblock \emph{Journal of School Violence}, 20(2): 167--181.

\bibitem[{Hanschmidt et~al.(2016)Hanschmidt, Lehnig, Riedel-Heller, and Kersting}]{hanschmidt2016stigma}
Hanschmidt, F.; Lehnig, F.; Riedel-Heller, S.~G.; and Kersting, A. 2016.
\newblock The stigma of suicide survivorship and related consequences—A systematic review.
\newblock \emph{PloS one}, 11(9).

\bibitem[{Hatton, Clark, and Huber(2024)}]{hatton2024temporal}
Hatton, V.~R.; Clark, E.; and Huber, R.~S. 2024.
\newblock Temporal patterns in youth suicide deaths reported in the National Violent Death Reporting System.
\newblock \emph{Journal of Adolescent Health}, 74(5): 1049--1052.

\bibitem[{Health and Human~Services(2025)}]{nimh2025suicide}
Health; and Human~Services, U. D.~o. 2025.
\newblock Suicide.
\newblock Technical report, National Institute of Mental Health.

\bibitem[{Jaycox et~al.(2024)Jaycox, Murphy, Zehr, Pearson, and Avenevoli}]{jaycox2024social}
Jaycox, L.~H.; Murphy, E.~R.; Zehr, J.~L.; Pearson, J.~L.; and Avenevoli, S. 2024.
\newblock Social Media and Suicide Risk in Youth.
\newblock \emph{JAMA Network Open}, 7(10): e2441499--e2441499.

\bibitem[{Jiang et~al.(2023)Jiang, Sablayrolles, Mensch, Bamford, Chaplot, de~las Casas, Bressand, Lengyel, Lample, Saulnier, Lavaud, Lachaux, Stock, Scao, Lavril, Wang, Lacroix, and Sayed}]{jiang2023mistral7b}
Jiang, A.~Q.; Sablayrolles, A.; Mensch, A.; Bamford, C.; Chaplot, D.~S.; de~las Casas, D.; Bressand, F.; Lengyel, G.; Lample, G.; Saulnier, L.; Lavaud, L.~R.; Lachaux, M.-A.; Stock, P.; Scao, T.~L.; Lavril, T.; Wang, T.; Lacroix, T.; and Sayed, W.~E. 2023.
\newblock Mistral 7B.
\newblock arXiv:2310.06825.

\bibitem[{John et~al.(2018)John, Glendenning, Marchant, Montgomery, Stewart, Wood, Lloyd, Hawton et~al.}]{john2018self}
John, A.; Glendenning, A.~C.; Marchant, A.; Montgomery, P.; Stewart, A.; Wood, S.; Lloyd, K.; Hawton, K.; et~al. 2018.
\newblock Self-harm, suicidal behaviours, and cyberbullying in children and young people: Systematic review.
\newblock \emph{Journal of medical internet research}, 20(4): e9044.

\bibitem[{Kavuluru et~al.(2016)Kavuluru, Ramos-Morales, Holaday, Williams, Haye, and Cerel}]{kavuluru2016classification}
Kavuluru, R.; Ramos-Morales, M.; Holaday, T.; Williams, A.~G.; Haye, L.; and Cerel, J. 2016.
\newblock Classification of helpful comments on online suicide watch forums.
\newblock In \emph{Proceedings of the 7th ACM international conference on bioinformatics, computational biology, and health informatics}, 32--40.

\bibitem[{Kumar et~al.(2015)Kumar, Dredze, Coppersmith, and De~Choudhury}]{kumar2015detecting}
Kumar, M.; Dredze, M.; Coppersmith, G.; and De~Choudhury, M. 2015.
\newblock Detecting changes in suicide content manifested in social media following celebrity suicides.
\newblock In \emph{Proceedings of the 26th ACM conference on Hypertext \& Social Media}, 85--94.

\bibitem[{Livingstone, Stoilova, and Kelly(2016)}]{livingstone2016cyberbullying}
Livingstone, S.; Stoilova, M.; and Kelly, A. 2016.
\newblock Cyberbullying: Incidence, trends and consequences.
\newblock \emph{Ending the torment: Tackling bullying from the schoolyard to cyberspace}.

\bibitem[{Macrynikola et~al.(2021)Macrynikola, Auad, Menjivar, and Miranda}]{macrynikola2021does}
Macrynikola, N.; Auad, E.; Menjivar, J.; and Miranda, R. 2021.
\newblock Does social media use confer suicide risk? A systematic review of the evidence.
\newblock \emph{Computers in Human Behavior Reports}, 3: 100094.

\bibitem[{Marciano et~al.(2022)Marciano, Ostroumova, Schulz, and Camerini}]{marciano2022digital}
Marciano, L.; Ostroumova, M.; Schulz, P.~J.; and Camerini, A.-L. 2022.
\newblock Digital media use and adolescents' mental health during the COVID-19 pandemic: a systematic review and meta-analysis.
\newblock \emph{Frontiers in public health}, 9: 793868.

\bibitem[{McCarthy(2010)}]{mccarthy2010internet}
McCarthy, M.~J. 2010.
\newblock Internet monitoring of suicide risk in the population.
\newblock \emph{Journal of affective disorders}, 122(3).

\bibitem[{Mezuk et~al.(2021)Mezuk, Kalesnikava, Kim, Ko, and Collins}]{mezuk2021not}
Mezuk, B.; Kalesnikava, V.~A.; Kim, J.; Ko, T.~M.; and Collins, C. 2021.
\newblock Not discussed: Inequalities in narrative text data for suicide deaths in the National Violent Death Reporting System.
\newblock \emph{PLoS one}, 16(7): e0254417.

\bibitem[{Mok, Jorm, and Pirkis(2015)}]{mok2015suicide}
Mok, K.; Jorm, A.~F.; and Pirkis, J. 2015.
\newblock Suicide-related Internet use: A review.
\newblock \emph{Australian \& New Zealand Journal of Psychiatry}, 49(8): 697--705.

\bibitem[{Nazarov et~al.(2019)Nazarov, Guan, Chihuri, and Li}]{nazarov2019research}
Nazarov, O.; Guan, J.; Chihuri, S.; and Li, G. 2019.
\newblock Research utility of the National Violent Death Reporting System: a scoping review.
\newblock \emph{Injury epidemiology}, 6: 1--12.

\bibitem[{NCES(2024)}]{edUSEducation}
NCES. 2024.
\newblock {U}.{S}. {E}ducation in the {T}ime of {C}{O}{V}{I}{D}.
\newblock \url{https://nces.ed.gov/surveys/annualreports/topical-studies/covid/}.
\newblock [Accessed 16-01-2025].

\bibitem[{O'Connor and Kirtley(2018)}]{o2018integrated}
O'Connor, R.~C.; and Kirtley, O.~J. 2018.
\newblock The integrated motivational--volitional model of suicidal behaviour.
\newblock \emph{Philosophical Transactions of the Royal Society B: Biological Sciences}, 373(1754): 20170268.

\bibitem[{Pantic(2014)}]{pantic2014online}
Pantic, I. 2014.
\newblock Online social networking and mental health.
\newblock \emph{Cyberpsychology, Behavior, and Social Networking}, 17(10).

\bibitem[{Patel et~al.(2024)Patel, Sumner, Bowen, Zwald, Yard, Wang, Law, Holland, Nguyen, Mower et~al.}]{patel2024predicting}
Patel, D.; Sumner, S.~A.; Bowen, D.; Zwald, M.; Yard, E.; Wang, J.; Law, R.; Holland, K.; Nguyen, T.; Mower, G.; et~al. 2024.
\newblock Predicting state level suicide fatalities in the united states with realtime data and machine learning.
\newblock \emph{npj mental health research}, 3(1): 3.

\bibitem[{Pourmand et~al.(2019)Pourmand, Roberson, Caggiula, Monsalve, Rahimi, and Torres-Llenza}]{pourmand2019social}
Pourmand, A.; Roberson, J.; Caggiula, A.; Monsalve, N.; Rahimi, M.; and Torres-Llenza, V. 2019.
\newblock Social media and suicide: a review of technology-based epidemiology and risk assessment.
\newblock \emph{Telemedicine and e-Health}, 25(10): 880--888.

\bibitem[{Rothwell(2023)}]{rothwell2023parenting}
Rothwell, J. 2023.
\newblock How parenting and self-control mediate the link between social media use and youth mental health.
\newblock \emph{Institute for Family Studies and Gallup}.

\bibitem[{Sawhney et~al.(2021)Sawhney, Joshi, Nobles, and Shah}]{sawhney2021towards}
Sawhney, R.; Joshi, H.; Nobles, A.; and Shah, R.~R. 2021.
\newblock Towards Emotion-and Time-Aware Classification of Tweets to Assist Human Moderation for Suicide Prevention.
\newblock In \emph{Proceedings of the International AAAI Conference on Web and Social Media}, volume~15, 609--620.

\bibitem[{Schulhoff et~al.(2024)Schulhoff, Ilie, Balepur, Kahadze, Liu, Si, Li, Gupta, Han, Schulhoff et~al.}]{schulhoff2024prompt}
Schulhoff, S.; Ilie, M.; Balepur, N.; Kahadze, K.; Liu, A.; Si, C.; Li, Y.; Gupta, A.; Han, H.; Schulhoff, S.; et~al. 2024.
\newblock The Prompt Report: A Systematic Survey of Prompting Techniques.
\newblock \emph{arXiv preprint arXiv:2406.06608}.

\bibitem[{Shain(2018)}]{shain2018youth}
Shain, B.~N. 2018.
\newblock Youth Suicide: The First Suicide Attempt.
\newblock \emph{Journal of the American Academy of Child and Adolescent Psychiatry}, 57(10): 730--732.

\bibitem[{Skaik and Inkpen(2020)}]{skaik2020using}
Skaik, R.; and Inkpen, D. 2020.
\newblock Using social media for mental health surveillance: a review.
\newblock \emph{ACM Computing Surveys (CSUR)}, 53(6): 1--31.

\bibitem[{Thulin et~al.(2024{\natexlab{a}})Thulin, French, Messman, Masi, and Heinze}]{thulin2024firearm}
Thulin, E.~J.; French, A.; Messman, E.; Masi, R.; and Heinze, J.~E. 2024{\natexlab{a}}.
\newblock Firearm-related tips in a statewide school anonymous reporting system.
\newblock \emph{Pediatrics}, 153(2).

\bibitem[{Thulin et~al.(2024{\natexlab{b}})Thulin, Kusunoki, Kernsmith, Smith-Darden, Grogan-Kaylor, Zimmerman, and Heinze}]{thulin2024longitudinal}
Thulin, E.~J.; Kusunoki, Y.; Kernsmith, P.~D.; Smith-Darden, J.~P.; Grogan-Kaylor, A.; Zimmerman, M.; and Heinze, J.~E. 2024{\natexlab{b}}.
\newblock Longitudinal effects of electronic dating violence on depressive symptoms and delinquent behaviors across adolescence.
\newblock \emph{Journal of interpersonal violence}, 39(11-12): 2526--2551.

\bibitem[{Twenge and Martin(2020)}]{twenge2020gender}
Twenge, J.~M.; and Martin, G.~N. 2020.
\newblock Gender differences in associations between digital media use and psychological well-being: Evidence from three large datasets.
\newblock \emph{Journal of adolescence}, 79: 91--102.

\bibitem[{Yang et~al.(2024)Yang, Yang, Hui, Zheng, et~al., and Fan}]{qwen2}
Yang, A.; Yang, B.; Hui, B.; Zheng, B.; et~al.; and Fan, Z. 2024.
\newblock Qwen2 Technical Report.
\newblock \emph{arXiv preprint arXiv:2407.10671}.

\bibitem[{Y{\i}ld{\i}z et~al.(2019)Y{\i}ld{\i}z, Orak, Walker, and Solakoglu}]{yildiz2019suicide}
Y{\i}ld{\i}z, M.; Orak, U.; Walker, M.~H.; and Solakoglu, O. 2019.
\newblock Suicide contagion, gender, and suicide attempts among adolescents.
\newblock \emph{Death studies}.

\bibitem[{Zsila and Reyes(2023)}]{zsila2023pros}
Zsila, {\'A}.; and Reyes, M. E.~S. 2023.
\newblock Pros \& cons: impacts of social media on mental health.
\newblock \emph{BMC psychology}, 11(1).

\end{thebibliography}

\section{Paper Checklist}

\begin{enumerate}

\item For most authors...
\begin{enumerate}
    \item  Would answering this research question advance science without violating social contracts, such as violating privacy norms, perpetuating unfair profiling, exacerbating the socio-economic divide, or implying disrespect to societies or cultures?
    \answerYes{Yes. NVDRS is a population-scale registry of suicide decedents in the USA. While it contains some information that could be identifying (e.g., demographics, county, and date of death), it does not have any personally identifying information. We have access NVDRS through a restricted access data agreement and our use of the data is in full compliance with this agreement. We have discussed demographic discrepancies in this dataset in the paper (section: Data).}
    
  \item Do your main claims in the abstract and introduction accurately reflect the paper's contributions and scope?
    \answerYes{Yes.}
   \item Do you clarify how the proposed methodological approach is appropriate for the claims made? 
    \answerYes{Yes, in each section.}
   \item Do you clarify what are possible artifacts in the data used, given population-specific distributions?
    \answerYes{Yes, see section Data.}
  \item Did you describe the limitations of your work?
   \answerYes{Yes, in each section.} 
   
  \item Did you discuss any potential negative societal impacts of your work?
    \answerYes{Yes, see section Data and Discussion.}
\item Did you discuss any potential misuse of your work?
    \answerYes{Yes, see section Data and Discussion.}
    \item Did you describe steps taken to prevent or mitigate potential negative outcomes of the research, such as data and model documentation, data anonymization, responsible release, access control, and the reproducibility of findings?
    \answerYes{Yes, see section Data and Discussion.}
  \item Have you read the ethics review guidelines and ensured that your paper conforms to them?
    \answerYes{Yes}
\end{enumerate}

\item Additionally, if your study involves hypotheses testing...
\begin{enumerate}
  \item Did you clearly state the assumptions underlying all theoretical results?
  \answerYes{Yes, see sections Temporal Variation and Heterogeneity by Decedent Characteristics}
  \item Have you provided justifications for all theoretical results?
  \answerYes{Yes, see sections Temporal Variation and Heterogeneity by Decedent Characteristics}
  \item Did you discuss competing hypotheses or theories that might challenge or complement your theoretical results?
  \answerYes{Yes, see sections Temporal Variation and Heterogeneity by Decedent Characteristics}
  \item Have you considered alternative mechanisms or explanations that might account for the same outcomes observed in your study?
  \answerYes{Yes, see sections Temporal Variation and Heterogeneity by Decedent Characteristics}
  \item Did you address potential biases or limitations in your theoretical framework?
  \answerYes{Yes, see section Data}
  \item Have you related your theoretical results to the existing literature in social science?
  \answerYes{Yes, see sections Online Activity and Youth Suicide, Thematic Analysis of Online Activities, Temporal Variation, and Heterogeneity by Decedent Characteristics}
  \item Did you discuss the implications of your theoretical results for policy, practice, or further research in the social science domain?
  \answerYes{Yes, see sections Temporal Variation, Heterogeneity by Decedent Characteristics, and Discussion}
\end{enumerate}

\item Additionally, if you are including theoretical proofs...
\begin{enumerate}
  \item Did you state the full set of assumptions of all theoretical results?
   \answerNA{}
   
  \item Did you include complete proofs of all theoretical results?
  \answerNA{}
    
\end{enumerate}

\item Additionally, if you ran machine learning experiments...
\begin{enumerate}
  \item Did you include the code, data, and instructions needed to reproduce the main experimental results (either in the supplemental material or as a URL)?
   \answerNo{No, because the data is restricted access so we can't share it, and we will provide a Github URL with the camera-ready submission.}
  \item Did you specify all the training details (e.g., data splits, hyperparameters, how they were chosen)?
  \answerYes{Yes}
 \item Did you report error bars (e.g., with respect to the random seed after running experiments multiple times)?
  \answerNo{No, because we used zero-shot learning with a temperature of 0 for reproducibility.}
\item Did you include the total amount of compute and the type of resources used (e.g., type of GPUs, internal cluster, or cloud provider)?
  \answerYes{Yes}
 \item Do you justify how the proposed evaluation is sufficient and appropriate to the claims made? 
  \answerYes{Yes}
 \item Do you discuss what is ``the cost`` of misclassification and fault (in)tolerance?
 \answerYes{Yes}
  
\end{enumerate}

\item Additionally, if you are using existing assets (e.g., code, data, models) or curating/releasing new assets, \textbf{without compromising anonymity}...
\begin{enumerate}
  \item If your work uses existing assets, did you cite the creators?
    \answerYes{Yes, see section Data}
  \item Did you mention the license of the assets?
    \answerYes{Yes, see section Data}
  \item Did you include any new assets in the supplemental material or as a URL?
    \answerNo{No, because the data is restricted access.}
  \item Did you discuss whether and how consent was obtained from people whose data you're using/curating?
    \answerNA{NA, not curating data}
  \item Did you discuss whether the data you are using/curating contains personally identifiable information or offensive content?
    \answerYes{Yes, see content warning}
\item If you are curating or releasing new datasets, did you discuss how you intend to make your datasets FAIR? 
    \answerNA{NA, not curating data}
\item If you are curating or releasing new datasets, did you create a Datasheet for the Dataset (see \citet{gebru2021datasheets})? 
    \answerNA{NA, not curating data}
\end{enumerate}

\item Additionally, if you used crowdsourcing or conducted research with human subjects, \textbf{without compromising anonymity}...
\begin{enumerate}
  \item Did you include the full text of instructions given to participants and screenshots?
    \answerNA{NA, all annotators were authors}
  \item Did you describe any potential participant risks, with mentions of Institutional Review Board (IRB) approvals?
    \answerNA{NA, all annotators were authors}
  \item Did you include the estimated hourly wage paid to participants and the total amount spent on participant compensation?
    \answerNA{NA, all annotators were authors}
   \item Did you discuss how data is stored, shared, and deidentified?
   \answerNA{NA, all annotators were authors}
\end{enumerate}

\end{enumerate}

\section{Ethics and Broader Impacts}

\revision{When working with machine learning and large datasets containing descriptive information about adolescent trauma and death, there are several ethical considerations and broader implications of the current findings. NVDRS utilizes police and medical examiner reports to create a publicly available dataset on all violent homicides, suicides, deaths caused by law enforcement, and unintentional firearm deaths. As NVDRS is a result of data extraction from reports of deceased individuals, there are no formal consent procedures from those who have died or from their next of kin. Police and medical examiner reports are largely treated as publicly available data, and NVDRS provides publicly available and restricted versions of data, which can be used to yield information that is helpful to decision makers, communities, and planners to understand, create interventions, and ultimately lower the risk of violent deaths. 

As described in the paper, there are known biases in narrative length that suggest that the quality of narratives and amount of information written about decedents is correlated with their demographic characteristics. For instance, male decedents and members of certain racial minorities tend to have shorter narratives. This is likely the result of biases in law enforcement investigations leading to less information being known about certain types of decedents. Findings need to be interpreted with these limitations in mind. We have attempted to center this point in our description of the data in the main paper. We also account for whether law enforcement investigated a decedent’s technology in order to obtain information about their online activities, in order to control for potential biases in quality of investigation.

Although the data in NVDRS is publicly available, as noted the Health insurance Portability and Accountability Act (HIPAA), a decedent’s personally identifiable information (PII) continued to be protected by the HIPAA Privacy Rule for 50 years. Direct identifiers, such as the decedent’s name and identifying document numbers (e.g., driver’s license) are removed from NVDRS data prior to it being made publicly available. Additionally, NVDRS does not contain personal characteristic data such as photographs, fingerprints, handwriting, nor biometric identifiers like retina scans, voice signatures, or facial geometry. However, characteristics such as race, place of birth, place of death, and additional contextual factors presented in narratives could be linked to social or mass media in order to reverse engineer the identities of individuals. This potential for linkage and reverse engineering is a reason why established data-sharing policies by the overseeing federal agency (CDC) and Institutional Review Board oversight of projects are critical to guiding research using NVDRS narrative and categorical data. This includes restricted versions of data for publicly available data that require additional levels of oversight and data security. Researchers in academic sectors and beyond need to not only follow these guidelines, but advocate for their existence and enforcement to ensure that technological advancements paired with rich narrative data involving adolescents (alive or deceased) do not cause harm. 
}

\appendix

\pagebreak

\section{Prompts}

\subsection{System Prompt}

I am a researcher studying suicide risk factors. You are a helpful AI question answering assistant, who answers all my questions.

\subsection{Step 1 Prompt: Split Narrative into Sentences}

Split the following narrative into sentences. Format your output as a list of all sentences in the narrative.

\subsection{Step 2 Prompt: Detect Online Activities}

\textit{Note: A narrative has online activities if the answers to 1, 3, and 4 are all 'Yes'}

1. Does the following sentence mention technology? Answer Yes or No. This includes online spaces, online activities, social media, web searches, messaging, chat, email, SMS, texting, viewing or posting content, gaming, online schooling, distance learning, phones, computers, laptops, electronic devices, porn, cyberbullying, or other technology. Answer Yes or No with no explanation.
2. If yes, what nouns, verbs, or phrases in the sentence talk about technology? Read the full sentence and provide a comma-separated list of all nouns, verbs, and phrases exactly as they appear in the text.
3. Do any of the activities involving technology require internet or SMS? Answer Yes or No. This includes having devices taken away. This does not include phone calls, voicemails, and passwords. This does not include mentions of phones or computers unless they also mention an activity that involves internet.
4. Did V (victim) participate in the activities involving technology? Answer Yes, No, or Unknown. 

You reply in JSON format with the fields 'technology', 'phrases', 'internet', 'participate'. If the answer to \#1 is No, 'phrases' should be an empty list and 'internet' and 'participate' should be No.  

Example Sentence: V is a 24-year-old male who was found with a gunshot wound to the head.
Example Answer: {'technology':'No', 'phrases': [], 'internet':'No', 'participate':'No'}

Example Sentence: His mother called 911 and started CPR.
Example Answer: {'technology':'Yes', 'phrases': ['called'], 'internet':'No', 'participate':'No'}

Example Sentence: The V's ex-boyfriend called 911 after his mom sent him text messages stating that V had a gun and was going to kill herself.
Example Answer: {'technology':'Yes', 'phrases': ['called', 'text messages'], 'internet':'Yes', 'participate':'No'}

Example Sentence: On review of the V's phone officers found numerous websites and chat forums containing suicidal discussions.
Example Answer: {'technology':'Yes', 'phrases': ['phone', 'websites', 'chat forums'], 'internet':'Yes', 'participate':'Yes'}

\subsection{Step 3 Prompt: Combine Related Sentences}

Does some part of the second sentence reference the first sentence? Answer Yes or No with no explanation.

First Sentence: {first}

Second Sentence: {second}

\subsection{Step 4 Prompt a: Detect Harm-Related Themes}

\textit{Note: If the theme is coded as Disclosure if option 'A' is selected; as Graphic Disclosure if on of ['B','C','D'] is selected; as Other-Harm Content if one of ['P','Q'] is selected; as Perpetrator if one of ['F','K','L'] is selected; as Self-Harm Content if one of ['E','M','N','O'] is selected; and as  Victim if one of ['I','J'] is selected.}

In the following sentence, which of the following is true? Select only one option. Give only the letter with no explanation.

A. V posted on social media, texted, or messaged someone indicating they were thinking about suicide, planning to kill or hurt themselves, or threatened to kill themselves

B. V posted a video, photo, or audio of their own suicide attempt

C. V posted a video, photo, or audio showing themselves with a weapon or showing their means of suicide

D. V was live-streaming or broadcasting their own suicide 

E. V searched how to kill or hurt themselves online, or ordered their means of death online

F. V threatened someone else (do not count: threatened to hurt or kill themselves)

G. V argued with someone

H. V talked about non-suicidal self-harm online

I. V was bullied, harassed, threatened, or harmed online

J. Someone was texting or posting inappropriate or harmful content about V (V was the victim)

K. V harmed someone else online; this includes bullying, threats, inappropriate behavior, explicit activities, hurting someone, etc. (V was the perpetrator)

L. V engaged in illegal activities online; this includes buying drugs, soliciting illicit photos, etc.

M. V was on a forum for suicide or self harm

N. V viewed other suicidal content online

O. V viewed other content about weapons or potential means of death online

P. V viewed other violent content online

Q. V viewed porn or other illegal or explicit content online

R. V was using their computer

S. V texted someone about something else

T. None of the above

\subsection{Step 4 Prompt b: Detect Interpersonal-Related Themes}

\textit{Note: If the theme is coded as Sharing if one of ['H','J','K'] is selected; as Online Conflict if one of ['E','F','G'] is selected; and as Online Relationship if one of ['N','O'] is selected.}

In the following sentence, which of the following is true? Select only one option. Give only the letter with no explanation.

A. V posted on social media, texted, or messaged someone indicating they were thinking about suicide or planning to kill or hurt themselves

B. V posted on social media, texted, or messaged someone indicating they were not planning to be around (e.g., 'goodbye') OR finality (e.g., 'i love you')

C. V posted a photo or video of their suicide online

D. V searched how to kill or hurt themselves online

E. V argued about something that happened online

F. V argued with someone online

G. V broke up or fought with a significant other via text or chat

H. V wrote about non-suicidal self-harm

I. V texted that they loved someone soon before their death

J. V wrote about their feelings

K. V shared or revealed something about their life

L. V texted a significant other, friend, or parent

M. V texted or posted something else

N. V was in an online relationship

O. V was talking to someone who they primarily knew online

P. V was being bullied, threatened, or harmed online

Q. V was bullying, threatening, or harming someone else online

R. V was texting someone

S. None of the above

\subsection{Step 4 Prompt c: Detect Activity Level and Misc Themes}

\textit{Note: If the theme is coded as Withdraw if one of ['J','O','P'] is selected; as Intensify if one of ['B','C','D','E','F'] is selected; and as School if 'I' is selected.}

In the following sentence, which of the following is true? Select only one option. Give only the letter with no explanation.

A. V posted on social media or messaged someone indicating they were thinking about suicide or planning to kill or hurt themselves

B. The sentence explicitly says that V was addicted to a website, app, or online platform

C. The sentence explicitly says that V was spending a large amount of time or too much time playing video games, on social media, or other online activities

D. The sentence explicitly says that V spent most of their time or only wanted to play video games or do other online activities

E. V had a high level of online activity, including sending a very high number of text messages, doing a large amount of online shopping, etc.

F. V wanted to spend time online instead of offline

G. V did not have their phone or computer, or V was not responding

H. V blocked or cut off contact with one person or a few people

I. V was doing online schooling and having problems or upset about it

J. V had stopped using social media, deleted an account, or withdrew from an online account

K. V wrote a suicide note, gave away their devices or passwords, or saw suicidal content online

L. Someone used V's phone or an app to find V

M. LE or police searched V's laptop, phone, or browser

N. V's laptop, phone, or browser were investigated or searched

O. V was not allowed to use their phone, computer, gaming, social media, or other electronic devices

P. Someone took away V's internet, phone, computer, gaming, social media, or other electronic devices

Q. V had online relationships or online friends

R. V searched the web

S. Someone found V by tracking location on their phone

T. None of the above apply, but V was playing video games, using their computer, texting, or doing some other online activity

U. None of the above

\subsection{Step 4 Prompt d: Detect Source of Information}

\textit{Note: If the source of information is coded as LE Searches `source' is `B'; as Next of Kin (Explicit) if `source' is `A'; as Next of Kin (Implicit) if `source\_known' is False and `sm' or `nok' is True; and Unknown otherwise.}

In the following sentence:

1- Does it describe V posting on social media?

2- Does it describe contact between V and someone else (friend, family, significant other, etc.)?

3- Does it explicitly state who reported information about V's online activities? 

4- If yes, is the information reported by A) next of kin (friend, family, significant other, etc.) or B) searches of V's devices (laptop, phone, etc.) during the death investigation?

Format your response as a json with attributes `sm', `nok', `source\_known', and `source'.

\begin{table*}[t]
\centering

\begin{tabular}{p{0.75in} p{1.25in} | p{4.5in}}
\textbf{Category} & 
\textbf{Theme} & 
\textbf{Description}\\ \hline \hline
\textbf{Harm to Self} &
Graphic Disclosure & Decedent live-streams their suicide, or posts a photograph, video, or audio segment showing their means of death or suicide plan.\\
 & Disclosure & Decedent messages or posts online indicating suicidal ideation or plans. This includes messages expressing some finality (e.g., `goodbye' or `I love you').\\
 & Self-Harm Content & Decedent searches, views, or shares content about suicide or lethal means.\\ \hline
\textbf{Harm to Others} &
Perpetrator & Decedent is the perpetrator of some type of harm (physical, psychological, legal, etc.) against another person.\\
 & Victim & Decedent is the victim of some type of harm, done by another person. \\
 & Other-Harm Content & Decedent views content online that involves some type of harm (physical, psychological, legal, etc.) to another real or fictional person.\\ \hline
\textbf{Interpersonal} & 
Online Conflict & Decedent is involved in an interpersonal conflict that starts or progresses online, or whose trigger was an online activity.\\
 & Personal Sharing & Decedent shares personal information, emotions, or vulnerable information; may be considered normative in online spaces (e.g., due to context collapse).\\
 & Online Relationships & Decedent is involved in a relationship (not necessarily romantic) that is primarily online, has multiple interactions, and involves personal sharing.\\ \hline
\textbf{Activity Levels} &
Withdrawing or Low Use & Decedent was less engaged in online activities than usual, had abandoned their online accounts, or had devices taken away (e.g., as punishment).\\
 & Intensifying or High Use & The decedent was frequently using or using more of their online accounts, as reported by next of kin.\\ \hline
\textbf{Life Events} &
Problems with Online School & Decedent was experiencing problems related to online education, including lack of attendance, performance issues, or negative feelings.\\
\end{tabular}
\caption{Themes describing online activities by youth who died by suicide, as reported in NVDRS.}
\label{tab:themes}
\end{table*}

\begin{table*}

\begin{tabular}{l||c|c|c|c|c}
& \textbf{LE Searches} & 
\textbf{NOK } & 
\textbf{NOK } & 
\textbf{NOK } & \textbf{Unknown} \\
&  & 
\textbf{(explicit)} & 
\textbf{(implicit: conversation)} & 
\textbf{(implicit: social media)} &  \\ \hline \hline
\textbf{LE Searches} & 14 &  &  &  & 2 \\
\textbf{NOK (explicit)} &  & 26 &  &  & 1 \\
\textbf{NOK (implicit: conversation)} &  & 9 & 15 &  & 1 \\
\textbf{NOK (implicit: social media)} &  &  &  & 19 & \\
\textbf{Unknown} & 1 &  &  & & 11 \\ \hline \hline
\textbf{Precision} & 0.93 & 0.74 & 1 & 1 & 0.73 \\
\textbf{Recall} & 0.88 & 0.96 & 0.60 & 1 & 0.92 \\
\end{tabular}
\caption{Confusion matrix depicting the performance of the pipeline to detect the source of information used for a decedent's online activities. Results are based on 100 randomly sampled cases where decedents' online activities were mentioned. The vertical axis depicts labels by a human annotator (ground truth), while the horizonal axis depicts labels by the LLM (predictions). We also show the precision and recall of each category.}
\label{tab:source-performance}
\end{table*}

\section{Keyword Evaluation}
\label{appendix:eval}

\begin{figure*}
    \centering
    \includegraphics[width=\textwidth]{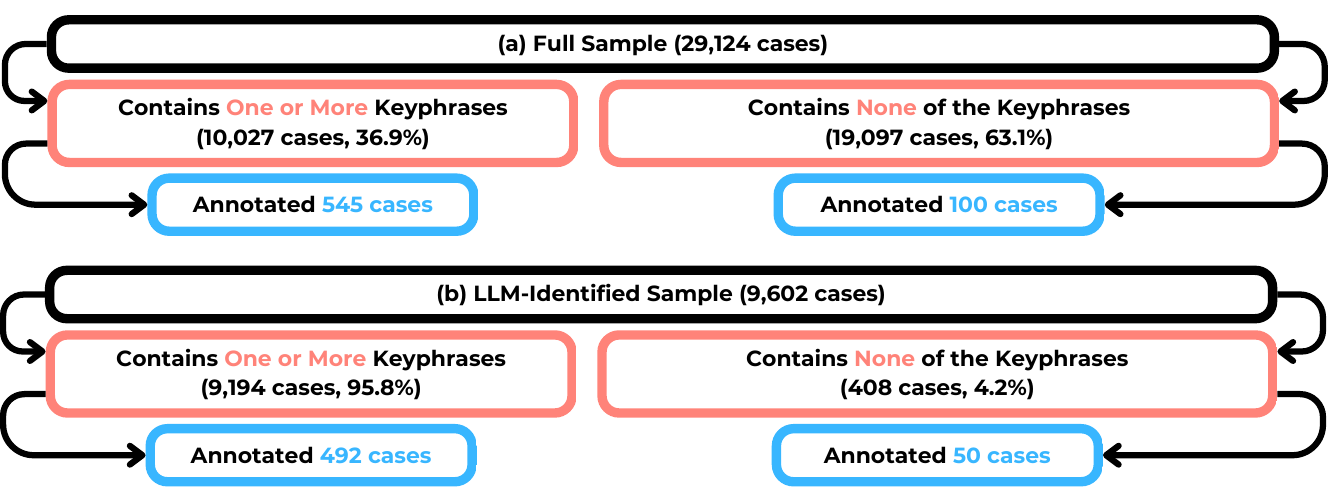}
    \caption{Cases for annotation were sampled using 51 keyphrases, allowing us to up-sample cases that mentioned online spaces without losing a lot of relevant cases (high recall). This flowchart shows how we sampled narratives to annotate a) during the thematic analysis (also used for model evaluation) and b) to get a conservative estimate of the recall of the keyphrases (in the LLM-identified sample).}
    \label{fig:flowchart}
\end{figure*}

\revision{Our goal is to use the 51 keyphrases to create a high recall subset of cases to facilitate a comprehensive annotation of cases related to online spaces. Using the 695 annotations we conducted (Figure~\ref{fig:flowchart}), we estimate precision and recall in two subsets of the data.

First, in the full dataset, we estimate that the recall of the keyphrases is near 1 and the precision is 0.804. We annotated 545 cases containing one or more keyphrases and 50 cases that did not contain any keyphrases. To calculate precision, we found that 438 of the 545 cases with keyphrases (80.4\%) actually referenced the decedent’s online activities. To calculate recall, we found that none of the 50 cases without any keyphrases pertained to online activities, giving a recall of 1 in our test set and suggesting near-perfect recall in the full sample. 

Second, we estimate a recall of 0.973 and a precision of 0.937 in a subset of the data that are likely relevant to the decedent’s online activities and, therefore, where the recall may be lower. To do this, we prompted Llama-3.1-8B with the prompt in Appendix Prompts to identify LLM-identified cases related to the decedent’s online activities, and calculated precision and recall using 591 annotated instances in the LLM-identified subset. To calculate precision, we found that 473 out of 505 (93.7\%) cases with keywords actually referenced the decedent’s online activities. To calculate recall, we found that 4.2\% of LLM-identified cases did not contain any of the 51 keyphrases. Of the 4.2\% of LLM-identified cases that didn’t contain our keyphrases, 60\% referenced a decedent’s online activities (30 out of 50 annotated cases), and of the 95.8\% of LLM-identified cases that do contain our keyphrases, 96.1\% (473 out of 492 annotated cases) pertain to online activities. Taking a weighted average of these two subsets gives a recall of 0.973 in the LLM-identified subset.
}

\begin{figure}
    \centering
    \includegraphics[width=\linewidth]{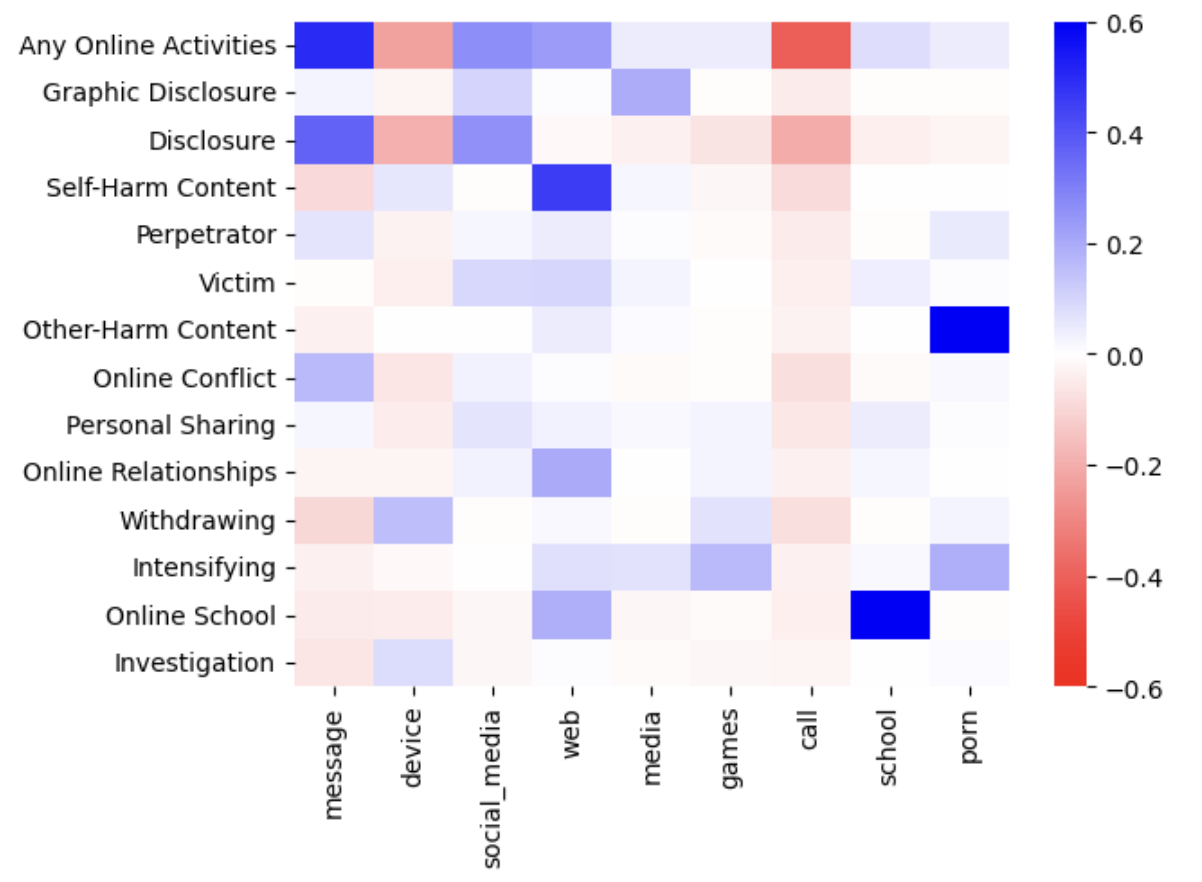}
    \caption{Correlation between the type of device mentioned in the narrative and the theme it's coded with.}
    \label{fig:types}
\end{figure}

\begin{figure}
    \centering
    \includegraphics[width=\linewidth]{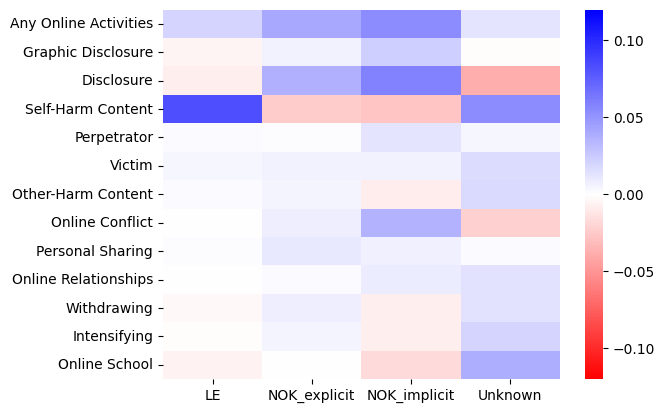}
    \caption{Correlation between the source of information mentioned in the narrative and the theme it's coded with.}
    \label{fig:sources}
\end{figure}

\begin{figure*}
    \centering
    \includegraphics[width = \textwidth]{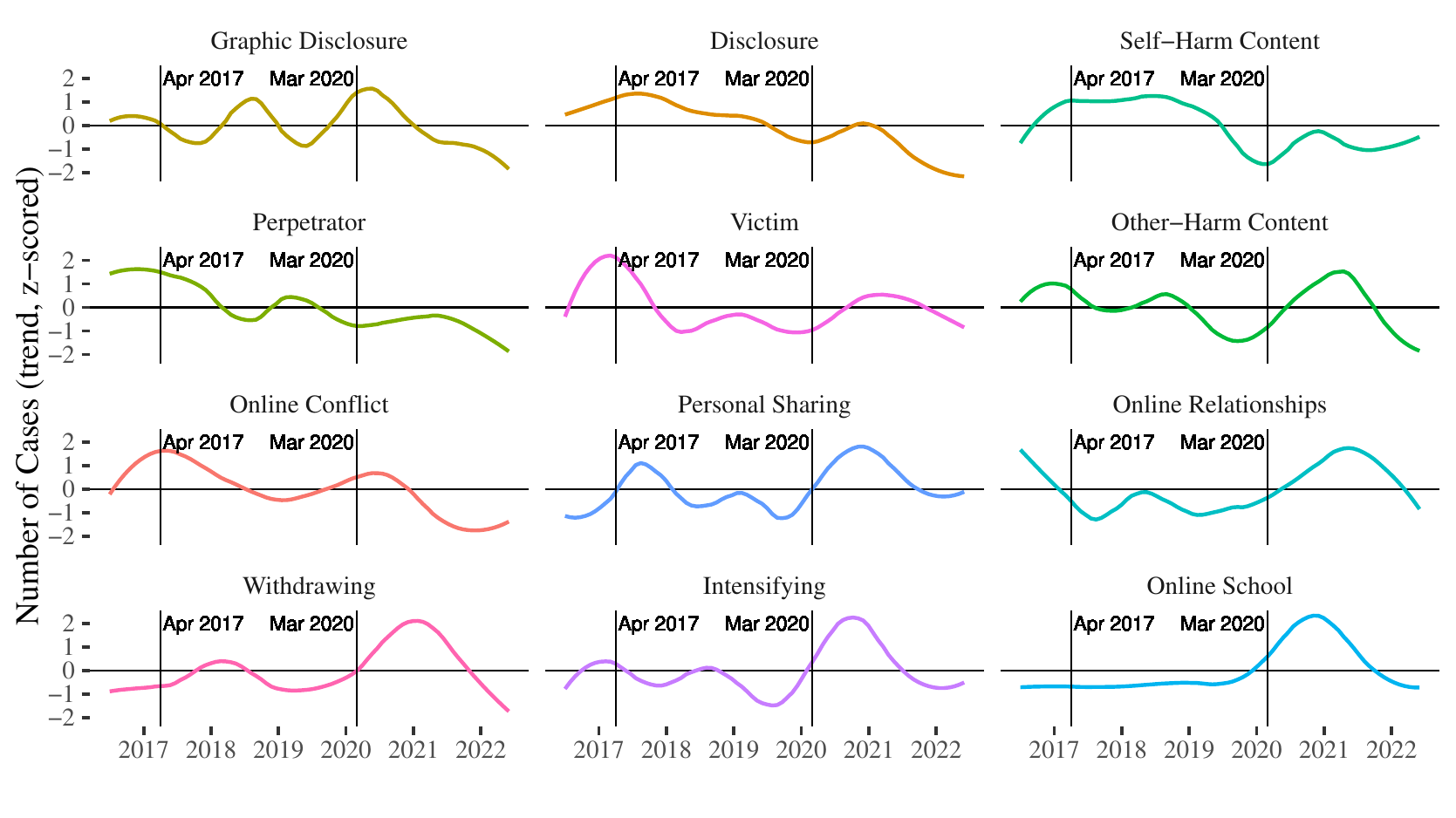}
    \caption{\revision{Temporal trend in the number of youth suicide deaths associated with each theme. For ease of comparison, we plot the z-scored trend of each time series.}}
    \label{fig:ts-all}
\end{figure*}

\begin{figure}
    \centering
    \includegraphics[width = \linewidth]{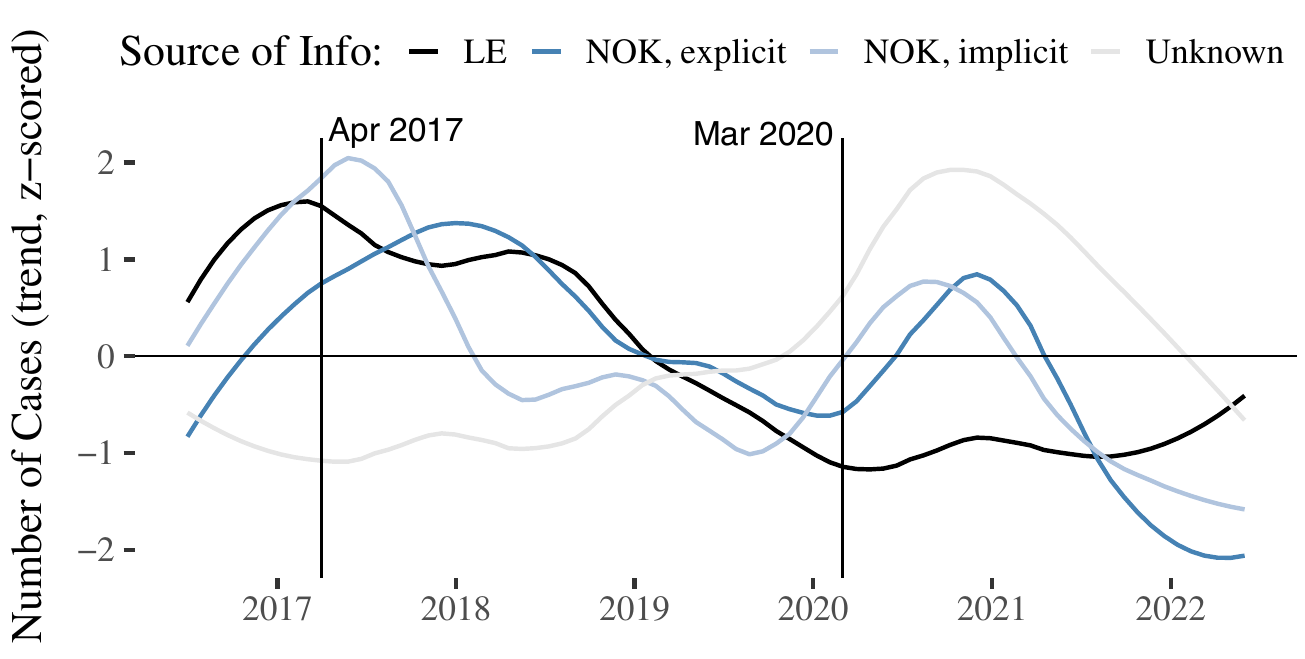}
    \caption{\revision{Temporal trend in the number of youth suicide deaths where information about the decedent's online activities came from each source of information: LE (LE searched the decedent's technology), NOK explicit (explicit references to next of kin reporting information about online activities), NOK implicit (the decedent communicated with next of kin directly or by posting on social media, implying that next of kin may have reported these interactions to LE/CME), Unknown (no explicit or implicit indicator of source of information). For ease of comparison, we plot the z-scored trend of each time series.}}
    \label{fig:ts-source}
\end{figure}

\begin{figure*}
    \centering
    \includegraphics[width=\textwidth]{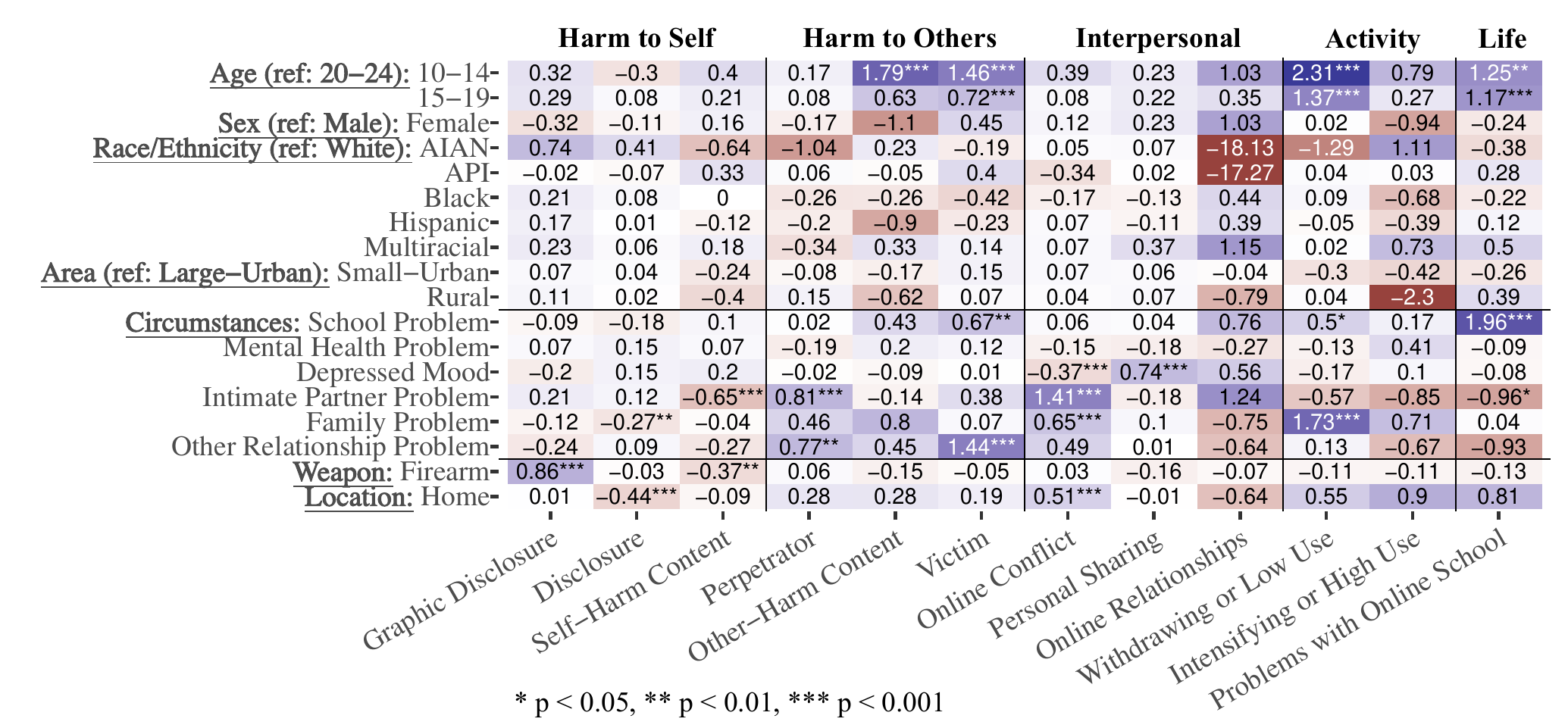}
    \caption{\revision{The association between each theme and the decedent's demographic characteristics and circumstances. For each theme, a logistic regression is used to calculate these associations using the presence of the theme as the dependent variable and the decedent characteristics and circumstances as independent variables. The heatmap visualizes the magnitude and direction of the regression coefficients, with darker blue representing larger positive coefficients, darker red representing larger negative coefficients, and white representing coefficients closer to 0. The regression coefficient is shown in each cell of the table. Asterisks represent the significance of each coefficient (* $p<0.05$, ** $p<0.01$, *** $p<0.001$), with p values adjusted for multiple comparisons using a Bonferroni correction.}}
    \label{fig:reg-table}
\end{figure*}

\end{document}